\documentclass[a4paper,twocolumn,11pt, accepted=2022-08-08]{quantumarticle}
\pdfoutput=1
\usepackage[utf8]{inputenc}
\usepackage[english]{babel}
\usepackage[T1]{fontenc}
\usepackage{amsmath, amsthm, amssymb,amsfonts,mathbbol,amstext}
\usepackage{hyperref}
\usepackage{tikz}
\usepackage{lipsum}
\usepackage{graphicx}
\usepackage{dcolumn}
\usepackage{bm}
\usepackage{mathtools}
\usepackage{color}
\usepackage{dsfont}
\usepackage[normalem]{ulem}
\usepackage{multirow}
\usepackage{diagbox}
\usepackage{bbm}
\usepackage{comment} 
\usepackage{marvosym}
\usepackage{xcolor}
\usepackage{appendix}
\usepackage[numbers,sort&compress]{natbib}

\newtheorem{theorem}{Theorem}
\newtheorem{result}[theorem]{Result}

\def\RR{\mathbbm{R}}

\begin{document}

\title{Events in quantum mechanics are maximally non-absolute}

\author{George Moreno}
\affiliation{International Institute of Physics, Federal University of Rio Grande do Norte, 59078-970, Natal, Brazil}
\affiliation{Departamento de Computação, Universidade Federal Rural de Pernambuco, 52171-900, Recife, Pernambuco, Brazil}
\email{marcosgeorge.mmf@gmail.com}
\author{Ranieri Nery}
\affiliation{International Institute of Physics, Federal University of Rio Grande do Norte, 59078-970, Natal, Brazil}
\email{ranieri.v.nery@gmail.com}
\author{Cristhiano Duarte}
\affiliation{International Institute of Physics, Federal University of Rio Grande do Norte, 59078-970, Natal, Brazil}
\affiliation{School of Physics and Astronomy, University of Leeds, Leeds LS2 9JT, United Kingdom}
\email{cristhianoduarte@gmail.com}
\author{Rafael Chaves}
\affiliation{International Institute of Physics, Federal University of Rio Grande do Norte, 59078-970, Natal, Brazil}
\affiliation{School of Science and Technology, Federal University of Rio Grande do Norte, Natal, Brazil}
\email{rafael.chaves@ufrn.br}

\date{August 16, 2022}

\maketitle

\begin{abstract}
The notorious quantum measurement problem brings out the difficulty to reconcile two quantum postulates: the unitary evolution of closed quantum systems and the wave-function collapse after a measurement. This problematics is particularly highlighted in the Wigner's friend thought experiment, where the mismatch between unitary evolution and measurement collapse leads to conflicting quantum descriptions for different observers. A recent no-go theorem has established that the (quantum) statistics arising from an extended Wigner's friend scenario is incompatible when one try to hold together three innocuous assumptions, namely no-superdeterminism, parameter independence and absoluteness of observed events. Building on this extended scenario, we introduce two novel measures of non-absoluteness of events. The first is based on the EPR2 decomposition, and the second involves the relaxation of the absoluteness hypothesis assumed in the aforementioned no-go theorem. To prove that quantum correlations can be maximally non-absolute according to both quantifiers, we show that chained Bell inequalities (and relaxations thereof) are also valid constraints for Wigner's experiment.
\end{abstract}

\section{Introduction}
In spite of the undeniable success of quantum mechanics as a physical theory, its foundations remain unsettled by a number of thorny issues. At the center of the debate is the measurement problem \cite{wigner1963problem,schlosshauer2005decoherence}.The measurement problem makes clear the apparent mismatch between the unitary, reversible and deterministic evolution of closed quantum systems and the non-unitary, irreversible, probabilistic behavior after a measurement in a collapse perspective. The problem arises from a reductionist perspective where macroscopic phenomena, like irreversible measurements and their consequent collapse, should arise from the behavior of its microscopic descriptions, like the unitary evolution described by the Schrödinger's equation \cite{Pusey18}.

The conceptual difficulties in the measurement problem become apparent in the gedanken experiment known as "Wigner's Friend" \cite{wigner1995remarks}. The scenario put forward by the thought experiment involves a quantum system in a superposition of states, an observer (Wigner's friend) performing measurements on it  and a super-observer (Wigner) who observes the friend and the friend's measurements on the quantum system. According to the standard collapse descriptions, Wigner's friend  describes the measurement as an irreversible process, by the collapse of the wave function describing the quantum system into one of the eigenstates of the observable being measured. In turn, Wigner might consider his friend and the system the friend is interacting with as a joint quantum system, so that from Wigner's perspective, the measurement process is described by a global unitary evolution generating an entangled state between both. In this case, Wigner traces out the state of his friend to obtain the final state of the system. Depending on who is observing, Wigner or Wigner's friend, we arrive then at two possible distinct descriptions for the same physical process, an apparent contradiction between the friend and Wigner's perspective, since the latter does not ascribe a well-defined value to the outcome associated with his friend’s well defined observation.

Eugene Wigner himself thought this gedanken experiment supported that consciousness was a necessary ingredient to describe quantum measurements \cite{wigner1995remarks}. Contrastly, Everett and his many-worlds interpretation \cite{everett2015relative}, reconciled the irreversible nature of measurements with an unitary behavior, since each of the possible outcomes observed by Wigner's friend would happen, only in different worlds. Resolutions based on hidden-variable models would required either faster-than-light signals \cite{bohm1966proposed}, superdeterminism \cite{hossenfelder2020rethinking,hooft2007free} or retrocausality \cite{price2008toy}.  To the Copenhagen \cite{stapp1972copenhagen}, Relational \cite{rovelli1996relational} or Quantum Bayesian interpretations \cite{caves2002quantum}, there is no issue brought about by the thought experiment, as the quantum state should be seen as relative to the observer -- a solution that might imply the rejection of the idea that measurement outcomes are absolute and agent-independent. Alternatively, one can introduce extra mechanisms in quantum theory, such as non-unitary quantum dynamics \cite{bassi2003dynamical} or the gravity-induced collapse of the wave function \cite{ghirardi1986unified,penrose1996gravity}, which would rule out the existence of macroscopic superpositions.

More recently, the interest in the measurement problem and its implications for the Wigner's friend experiment were reignited by new results \cite{Brukner2015,brukner2018no,Cavalcanti2021,frauchiger2018quantum,guerin2021no,healey2018quantum,proietti2019experimental,zukowski2021physics,cavalcanti2021view,Bong2020,xu2021nogo,Nurgalieva2018, Baumann2021} showing how the assumptions of the universality of quantum theory (its applicability to describe micro as well macroscopic phenomena) and the absoluteness of events (globally well-defined and observer-independent) imply experimentally testable constraints. Of particular relevance to us is the no-go theorem in Ref. \cite{Bong2020}, that building up on the earlier results of Brukner \cite{Brukner2015,brukner2018no} shows that there are quantum predictions that are incompatible with the conjunction of no-superdeterminism (NSD), parameter independence (PI) and absoluteness of observed events (AOE). Similar to Bell's theorem \cite{bell1964einstein}, such assumptions imply constraints on the statistics arising out of that scenario, called local friendliness (LF) inequalities in \cite{Bong2020}, that must be obeyed by any correlations compatible with them.

Leveraging the scenario established in Ref. \cite{Bong2020}, our aim in this paper is two-fold. First, we introduce two different measures of non-absoluteness of observed events. Second, we prove that quantum correlations can reach the maximum of such measures, thus proving that quantum events can be maximally non-absolute. The first measure we consider is based on the EPR2 decomposition \cite{elitzur1992quantum}, the fraction of events of a given observed probability distribution that cannot be considered to be absolute. The second measure is based on the relaxation of the assumption that events should be absolute. How much should we give up on absoluteness in order to explain quantum correlations? With that aim, we prove that the a relaxed version of the chained Bell inequalities \cite{braunstein1990wringing}
constrains the set of correlations compatible with absoluteness of events.

The paper is organized as follows. In Sec. \ref{sec:sec2} we review the EWFS introduced in \cite{Bong2020}. In Sec. \ref{sec:sec3} we introduce the two measures of non-absoluteness employed in the paper. In Sec. \ref{sec:sec4} we prove that the chained inequality is also a valid LF inequality and generalize it to the case where the AOE assumption is relaxed. Based on this inequality we prove our main result, that quantum correlations can lead to events that are maximally non-absolute. In Sec. \ref{sec:sec5} we discuss our findings and point out interesting future research directions.

\section{The extended Wigner's Friend Scenario}
\label{sec:sec2}

\begin{figure}
    \centering
    \includegraphics[scale = 0.35]{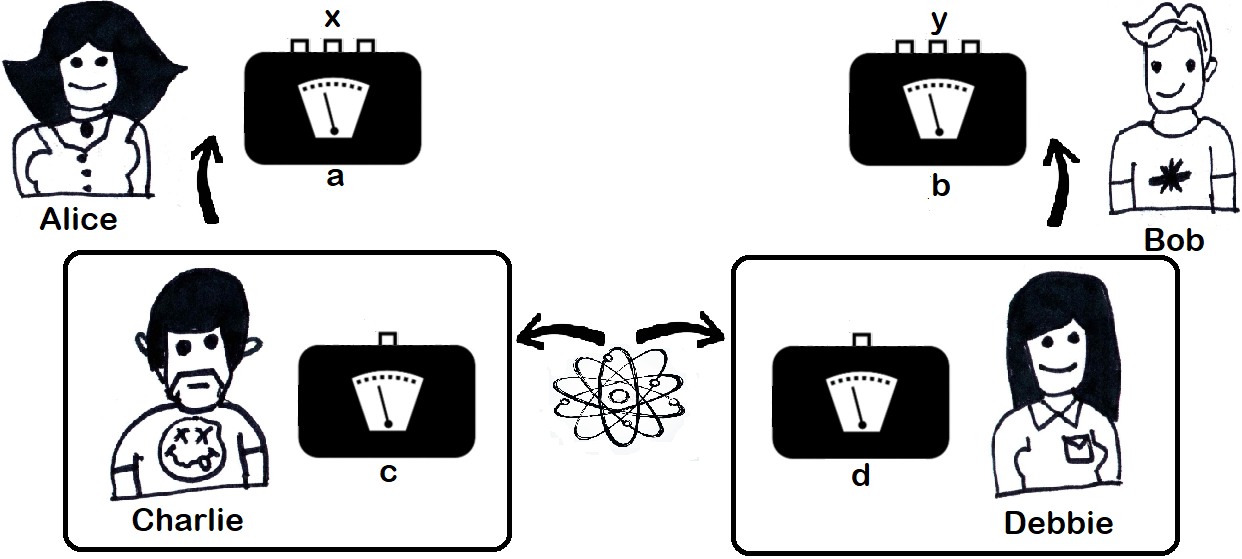}
    \caption{\textbf{Pictorial illustration of the extended Wigner's friend experiment}. The observers Charlie and Debbie measure their parts of an entangled system, obtaining measurement outcomes $c$ and $d$, respectively. In turn, the super-observers Alice and Bob input $x$ and $y$ (respectively) and then observe their friends and measurement devices, obtaining outcomes $a$ and $b$, respectively.}
    \label{fig:wigner}
\end{figure}

The bipartite extended Wigner's friend scenario (EWFS), as introduced in Ref. \cite{Bong2020} consists of two observers (the friends), Charlie and Debbie, and two superobservers Alice and Bob (the Wigners). Charlie and Debbie share a source of correlations on which they always perform some fixed measurement each inside a completely isolated laboratory, assumed to be space-like separated, obtaining the outcomes $c,\;d\in\{0,\dots,k-1\}$ for Charlie and Debbie respectively. Alice observes the laboratory containing Charlie and applies a measurement $x\in\{0,\dots,m-1\}$ obtaining the outcome $a\in\{0,1\}$ in the following way: if $x=m-1$ then $a=c$, which is equivalent to opening Charlie's laboratory and ask about the outcome Charlie has obtained; if $x\neq m-1$ Alice performs an arbitrary measurement on the laboratory. The same is done by Bob and Debbie. See Fig. \ref{fig:wigner} for an illustration of the EWFS.

To test the objectiveness of the friends's observations, that is, the fact that what the friend observes is "out there" and that its truth value is not subjective to the agent observing it, the EWFS relies on three assumptions, the conjunction of which the authors in \cite{Bong2020} call local friendliness (LF). The first of these assumptions, referred to as \emph{absoluteness of observed events} (AOE), states the existence of a joint distribution $\mathcal{P}(a,b,c,d|x,y)$ such that the experimentally obtained joint probability distribution $p(a,b|x,y)$ satisfies:
\begin{eqnarray}
\label{eq: AOE}
    \left\{\begin{array}{l}
         p(a,b|x,y)  =  \sum_{c,d}\mathcal{P}(a,b,c,d|x,y),\\
         p(a=c|x=m-1,y)  =  1,\\
         p(b=d|x,y=m-1)  =  1.
    \end{array}\right.
\end{eqnarray}
Notice that we choose to write this condition in a slightly different way than reference \cite{Bong2020}. This makes the problem more approachable from the numerical perspective while remaining completely equivalent to the original definition.

The AOE assumption assigns values only to observed outcomes. Since we deal with the probability distribution $p(a,b|x,y)$ observed in a given experiment realized by the two Wigners (Alice and Bob), we do not impose that measurements that are not performed do have results. This is quite different from the local hidden variable model entering in Bell's theorem, that according to Fine's theorem \cite{fine1982hidden} implies that even the outcomes of non-performed measurements should have a well defined probability distribution. In the EWFS, this is only the case for Alice's and Bob's outcome associated with measurements $x=m-1$ and $y=m-1$. More precisely, while a behavior $p(a,b|x,y)$ compatible with the Bell scenario can be associated with a distribution $p(a_0,a_1,\dots,a_{m - 1},b_0,b_1,\dots,b_{m - 1}$, a behavior $p(a,b|x,y)$ compatible with LF can only be associated with a distribution $p(a,b,a_{m - 1}, b_{m - 1}|x,y)$ or just $p(a,b,c,d|x,y)$. Under the AOE assumption, these special outcomes have values even when the corresponding measurements are not performed, that is, $x\neq m-1$ and $y \neq m-1$. That follows from the fact that the outcome of Alice associated with the measurement $x=m-1$ is encoded in the outcome $c$, which is measured in every run of the experiment by Charlie (a similar argument holding for Bob and Debbie). 

The second assumption, is that of \emph{no-superdeterminism} (NSD), expressed as
\begin{equation}
\label{eq: NSD}
    p(c,d|x,y) = p(c,d).
\end{equation}
stating that $c$ and $d$ are independent of $x$ and $y$, or equivalently that the choices of Alice and Bob cannot affect the results of Charlie and Debbie and the correlations shared among them. Notice that this is reminiscent of the free-will or measurement independence assumption in Bell's theorem \cite{hall2010local,chaves2015unifying,hall2020measurement,chaves2021causal}.

The third assumption, which we call \emph{parameter independence} (PI) (named locality assumption in \cite{Bong2020}), states that
\begin{equation}
\label{eq: NS}
    \left\{\begin{array}{l}
         p(a|c,d,x,y) = p(a|c,d,x),\;\;\forall\; a,c,d,x,y,  \\
         p(b|c,d,x,y) = p(b|c,d,y),\;\;\forall\; b,c,d,x,y . 
    \end{array}\right.
\end{equation}
Notice that the conditions above imply the no-signalling constraints \cite{popescu1994quantum} over the observed probability distribution $p(a,b\vert x,y)$, that is,
\begin{equation}
\label{eq: NS2}
    \left\{\begin{array}{l}
         p(a|x,y) = p(a|x),\;\;\forall\; a,x,y,  \\
         p(b|x,y) = p(b|y),\;\;\forall\; b,x,y . 
    \end{array}\right.
\end{equation}
The conjunction of assumptions \eqref{eq: AOE}, \eqref{eq: NSD} and \eqref{eq: NS} define the set of \emph{local friendliness} (LF) correlations that we call $\mathcal{S}^{LF}$.

\section{Quantifying the absoluteness of events}
\label{sec:sec3}

It is clear that if we drop the key conditions $p(a|c,x=m-1,y) = \delta_{a,c}$ and $p(b|d,x,y=m-1) = \delta_{b,d}$ from the AOE assumption, then the set of correlations compatible with \eqref{eq: AOE}, \eqref{eq: NSD} and \eqref{eq: NS} is equal to the set of non-signalling correlations defined by \eqref{eq: NS2} and that we denote by $\mathcal{S}^{NS}$. From this observation we can introduce two natural measures quantifying the absoluteness of events.

The first one is inspired by the EPR2 decomposition \cite{elitzur1992quantum}, and quantifies the fraction of events associated with a given probability distribution that cannot be described by a local friendly model. Quite generally, as $\mathcal{S}^{LF}$ is a convex polytope contained in $\mathcal{S}^{NS}$, any observed distribution on the EWFS can be decomposed as
\begin{multline}
\label{eq:EPR2}
    p(a,b \vert x,y)=q\cdot p_{LF}(a,b \vert x,y) \\+(1-q) \cdot p_{NS}(a,b \vert x,y),
\end{multline}
where $p_{LF}(a,b | x,y) \in \mathcal{S}^{LF}$ and $p_{NS}(a,b \vert x,y) \in \mathcal{S}^{NS}$. In other words, for a certain probability distribution $p(a,b|x,y)$, the convex decomposition of eq.~\eqref{eq:EPR2} explores the fractions $q$ of local-friendliness and $1-q$ of non-signaling that $p(a,b|x,y)$ admits. Put another way, one may want to see a particular decomposition of $p(a,b | x,y)$ as in eq.~\eqref{eq:EPR2} as a signature of the fraction $q$ of local-friendliness an event might have. In this sense, our first measure of absoluteness of events, that we call \emph{non-absoluteness fraction} $\mathcal{A}_f$, is then defined as the minimal fraction of non-absolute events weight over all possible decompositions as in \eqref{eq:EPR2}, that is,
\begin{equation}
\label{eq:measure1}
    \mathcal{A}_f( p(a,b \vert x,y))= \min_{p_{LF},p_{NS}}(1-q),
\end{equation}
an optimization that can be performed using a linear programming \cite{fitzi2010non}.
Notice that LF correlations have $\mathcal{A}_f=0$ while correlations with maximally non-absolute events have $\mathcal{A}_f=1$. A possible interpretation of $\mathcal{A}_{f}$ can be given as follows. If one considers the that joint probabilities have a frequentist meaning, in a statistical experiment, repeated many times, the non-absoluteness fraction $\mathcal{A}_f$ can be understood as the fraction of runs of the experiment where the observed measurements outcomes can be deemed to be non-absolute, in the sense that they are incompatible with the conditions $p(a=c|x=m-1,y)=1$ and $p(b=d|x,y=m-1)=1$.

Notice that any LF inequality provides a lower bound to the non-absoluteness fraction. Given a generic LF inequality of the form $I(p(a,b\vert x,y))=\sum_{a,b,x,y} \omega_{a,b,x,y} p(a,b\vert x,y) \leq \Omega_{LF}$ where $\Omega_{LF}$ is the maximum value achievable by a LF correlation, it follows that 
\begin{equation}
\label{eq:af_bound}
\mathcal{A}_f \geq 1-\frac{\Omega_{NS}-\Omega_{Q}}{\Omega_{NS}-\Omega_{LF}},
\end{equation}
where $\Omega_{NS}$ is the maximum value achievable by no-signalling correlations and $\Omega_Q$ is the value corresponding to a given quantum probability distribution under test. This bound follows from the fact that $I(p)=\Omega_Q=q\cdot I(p_{LF})+(1-q) \cdot I(p_{NS}) \leq q \cdot \Omega_{LF}+(1-q) \cdot \Omega_{NS}$. In particular, notice that if $\Omega_{Q}=\Omega_{NS}$, that is, the quantum value achieves the maximum no-signalling violation of the LF inequality, then it follows that $\mathcal{A}_f=1$.

As mentioned, the mathematical expression of $\mathcal{A}_f$ is similar to the so-called non-local fraction $\mathcal{N}_f$, introduced in Ref. \cite{elitzur1992quantum}, that quantifies the non-locality of a given distribution $p(a,b \vert x,y)$ in a Bell scenario. The key difference, however, is the fact that in the Bell case the probability $p_{LF}$ should be substituted by $p_{Local}$, that is, by probabilities compatible with local hidden variable models. Since the local friendly set of correlations is in general bigger than the local set, we have that, for a given $p(a,b\vert x,y)$, $\mathcal{N}_f \geq \mathcal{A}_f $---recall that we are minimizing the function $(1-q)$. In particular, we can have situations where $\mathcal{N}_f >0$ but $\mathcal{A}_f=0$. Regarding the bound in Eq. \eqref{eq:af_bound} the same idea holds. In the case of a Bell scenario we would need to replace $\Omega_{LF}$ by $\Omega_{Local}$, the bound of a given inequality respected by Bell local correlations.

As a second measure of absoluteness of events we consider a relaxation of the condition in \eqref{eq: AOE}, that we call \emph{$\epsilon-$AOE}. Namely, we now allow that
\begin{eqnarray}
\label{eq: AOE relaxed}
\left\{\begin{array}{l}
          p(a,b|x,y) = \sum_{c,d}\mathcal{P}(a,b,c,d|x,y),\\
          p(a=c|x=m-1,y) \geq 1 - \epsilon, \\
          p(b=d|x,y=m-1) \geq 1 - \epsilon .\\
     \end{array}\right.
\end{eqnarray}

\begin{figure}
    \centering
    \includegraphics[scale = 0.45]{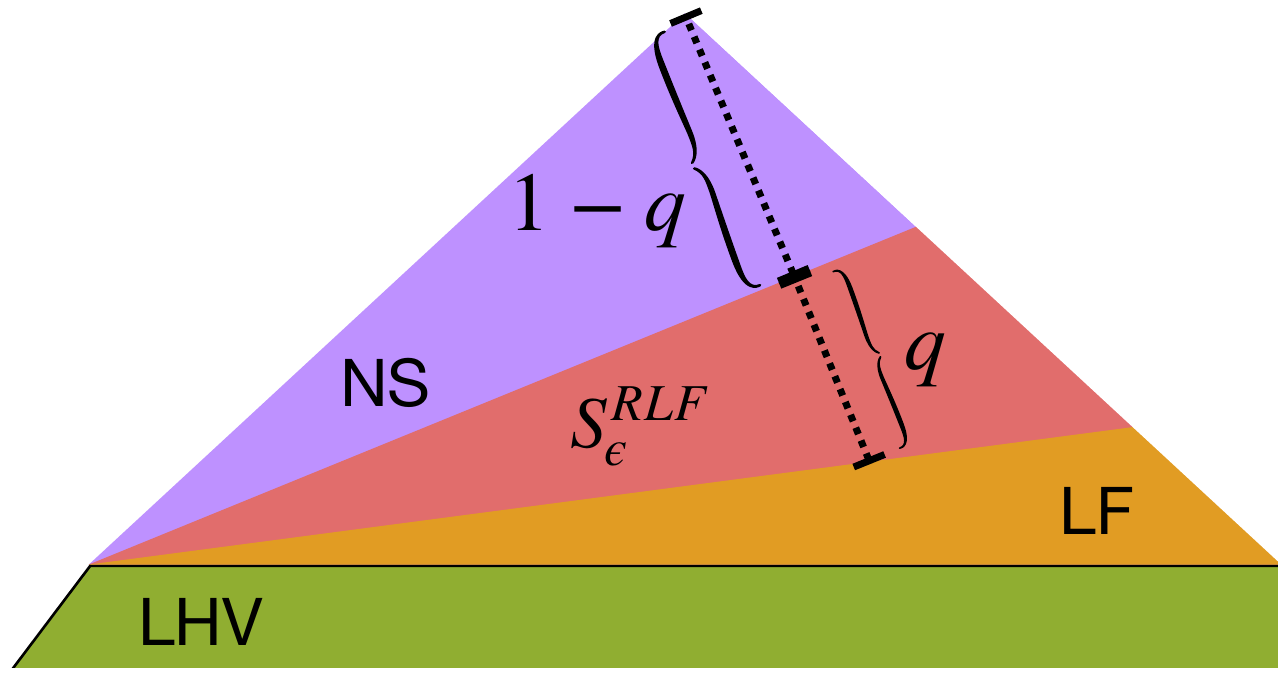}
    \caption{\textbf{Pictorial illustration of the $\mathcal{S}^{LF}$, its relaxed version  and $\mathcal{S}^{NS}$.} As stated in the main text $\mathcal{S}^{LF} \subseteq \mathcal{S}^{RLF}_{\epsilon} \subseteq \mathcal{S}^{RLF}_{\epsilon'} \subseteq \mathcal{S}^{NS}$ whenever $\epsilon\leq\epsilon'$. Furthermore $\mathcal{S}^{LHV} \subseteq \mathcal{S}^{LF}$ where $\mathcal{S}^{LHV}$ refers to the set of Bell local correlations. A point over the hypersurface limiting $\mathcal{S}^{RLF}_{\epsilon}$ has $\mathcal{A}_c=2\epsilon$ and $\mathcal{A}_f=(1-q)$.}
    \label{fig: absoluteness fraction}
\end{figure}

The conjunction of \eqref{eq: AOE relaxed}, \eqref{eq: NSD} and \eqref{eq: NS}  define  the set of \emph{relaxed local friendliness} (RLF) correlations denoted as $\mathcal{S}^{RLF}_{\epsilon}$. Clearly
\begin{equation}
    \mathcal{S}^{LF} \subseteq \mathcal{S}^{RLF}_{\epsilon} \subseteq \mathcal{S}^{RLF}_{\epsilon'} \subseteq \mathcal{S}^{NS}.
\end{equation}
whenever $\epsilon\leq\epsilon'$. See Fig. \ref{fig: absoluteness fraction} for a pictorial illustration.

From this we can define the \emph{non-absoluteness coefficient} $\mathcal{A}_c$, given by
\begin{equation}
\label{eq:measure2}
    \mathcal{A}_c( p(a,b \vert x,y))= \min 2\epsilon \text{  s.t.  } p(a,b \vert x,y) \in \mathcal{S}^{RLF}_{\epsilon}   .
\end{equation}
Observed distributions compatible with LF have $\mathcal{A}_c=0$ while maximally non-absolute events are those where $\mathcal{A}_c=1$, thus implying that the measurement outcome of Alice when her input is $x=m-1$ can be completely uncorrelated from that of her friend Charlie (similarly for Bob and Debbie).

Notice that the non-absoluteness coefficient $\mathcal{A}_c$, as opposed to the non-absolutness fraction $\mathcal{A}_f$, has no analogue in a Bell scenario, since it is defined via the relaxation of the absoluteness of events, an assumption that plays absolutely no role in Bell's theorem. In our view, it is precisely the relaxation of the absoluteness of events (AOE) that makes clear the distinction between the extended Wigner scenario (local friendliness) and the Bell scenario (local realism). As for the interpretation of the non-absolutness coefficient $\mathcal{A}_c$, notice that is does not take the conditions $p(a=c|x=m-1,y)=1$ and $p(b=d|x,y=m-1)=1$ for granted. Quite the contrary, asks how much such conditions have to be relaxed, in the form $p(a=c|x=m-1,y)\geq 1- \epsilon$ and $p(b=d|x,y=m-1)=1- \epsilon$, such that the observed data admits an explanation. In this sense, the measure $\mathcal{A}_c$ is related to a correlation set that contains the local friendly set and tends to the non-signalling set as $\mathcal{A}_c \rightarrow 1$.

Our first result shows the relation between both measures.
\begin{result}
The non-absoluteness coefficient is upper bounded by the non-absoluteness fraction, that is
\begin{equation}
\label{eq:relationAcAf}
\mathcal{A}_c \leq \mathcal{A}_f.
\end{equation}
\end{result}
\begin{proof}
First, we prove that $\mathcal{A}_c$ is a convex function: Let $p(a,b\vert x,y) = q\,p_A(a,b\vert x,y) + (1-q)\,p_B(a,b\vert x,y)$, with $0 \leq q \leq 1$. Assume that $\mathcal{A}_c(p_A) = 2\epsilon_A$ and $\mathcal{A}_c(p_B) = 2\epsilon_B$ and assume that these optimal values are attainable with global distributions $\mathcal{P}_A(a,b,c,d\vert x,y)$ and $\mathcal{P}_B(a,b,c,d\vert x,y)$, respectively. 

A possible attempt for solving $\mathcal{A}_c(p)$ is thus provided by $q\,\mathcal{P}_A(a,b,c,d\vert x,y) + (1-q)\,\mathcal{P}_B(a,b,c,d\vert x,y)$, which enables us to ensure that, at least, $p(a,b\vert x,y) \in \mathcal{S}^{RLF}_{\epsilon}$ for $\epsilon = q\,\epsilon_A + (1-q)\,\epsilon_B$. Since the proposed solution is not necessarily optimal, we must have $\mathcal{A}_c(p) \leq q\, 2\epsilon_a + (1-q)\,2\epsilon_B$, which, by definition, is the same as
\begin{equation}
\nonumber
\mathcal{A}_c(q\,p_A + (1-q)\,p_B) \leq q\,\mathcal{A}_c(p_A) + (1-q)\,\mathcal{A}_c(p_B).
\end{equation}

Consider now a possible realization for $\mathcal{A}_f(p)$, given by $p_{LF}$ and $p_{NS}$, i.e. $p = q^*\,p_{LF} + (1-q^*)\,p_{NS}$, where $1-q^* = \mathcal{A}_f(p)$. Using the convexity of $\mathcal{A}_c$ upon this decomposition leads to
\begin{equation}
\nonumber
    \mathcal{A}_c(p) \leq q^*\,\mathcal{A}_c(p_{LF}) + (1-q^*)\,\mathcal{A}_c(p_{NS}),
\end{equation}
but since $\mathcal{A}_c(p_{LF}) = 0$, we identify the rhs of the above expression with $\mathcal{A}_f(p)\,\mathcal{A}_c(p_{NS})$, leading to the inequality
\begin{equation}
\nonumber
    \mathcal{A}_c(p) \leq \mathcal{A}_f(p)\,\mathcal{A}_c(p_{NS}),
\end{equation}
for a given $p_{NS}$ participating in an optimal decomposition of $p(a,b\vert x,y)$ for $\mathcal{A}_f(p)$. We could tighten the inequality by performing a linear program minimizing $\mathcal{A}_c(p_{NS})$ within every possible decomposition for $p$ that has a fixed weight $1-q = \mathcal{A}_f(p)$. Going on the opposite direction, however, we may loosen the inequality by considering the general bound $\mathcal{A}_c(p_{NS}) \leq 1$, which leads to the general lower bound for $\mathcal{A}_f$: 
\begin{equation}
    \mathcal{A}_c(p) \leq \mathcal{A}_f(p).
\end{equation}

\end{proof}

To illustrate both measures of non-absoluteness in a concrete scenario, we first consider relaxations of the complete set of LF inequalities derived in \cite{Bong2020} for scenario where $m=3$ and $k=2$. This case is completely characterized by 6 classes of non-trivial LF inequalities given by
\begin{eqnarray}
\nonumber
I_1 & = & -\langle A_2\rangle - \langle A_1\rangle - \langle B_2\rangle - \langle B_1\rangle - \langle A_2B_2\rangle \\
\nonumber
& - & 2\langle A_2B_1\rangle - 2\langle A_1B_2\rangle + 2\langle A_1B_1\rangle - \langle A_1B_0\rangle\\
\nonumber
& - & \langle A_0B_1\rangle - \langle A_0B_0 \rangle \overset{\Omega_1}{\leq} 6,
\end{eqnarray}
\begin{eqnarray}
\nonumber
I_2 & = & -\langle A_2\rangle - \langle A_1\rangle - \langle A_0\rangle - \langle B_2\rangle - \langle A_2B_2\rangle \\
\nonumber
& - & \langle A_1B_2\rangle - \langle A_0B_2\rangle - 2\langle A_2B_1\rangle + \langle A_1B_1\rangle \\
\nonumber
& + & \langle A_0B_1\rangle - \langle A_1B_0\rangle +\langle A_0B_0\rangle \overset{\Omega_2}{\leq} 5,
\end{eqnarray}
\begin{eqnarray}
\nonumber
I_3 & = & -\langle A_2\rangle + \langle A_1\rangle + \langle B_2\rangle - \langle B_1\rangle + \langle A_2B_2\rangle \\
\nonumber
& - & \langle A_2B_1\rangle - \langle A_2B_0\rangle - \langle A_1B_2\rangle + \langle A_1B_1\rangle \\
\nonumber
& - & \langle A_1B_0\rangle - \langle A_0B_0\rangle -\langle A_0B_1\rangle \overset{\Omega_3}{\leq} 4,
\end{eqnarray}
\begin{eqnarray}
\nonumber
I_4 & = & -\langle A_1\rangle - \langle A_0\rangle - \langle B_1\rangle - \langle B_0\rangle - \langle A_2B_1\rangle \\
\nonumber
& + & \langle A_2B_0\rangle - \langle A_1B_2\rangle - \langle A_1B_1\rangle - \langle A_1B_0\rangle \\
\nonumber
& + & \langle A_0B_2\rangle - \langle A_0B_1\rangle -\langle A_0B_0\rangle \overset{\Omega_4}{\leq} 4,
\end{eqnarray}
\begin{eqnarray}
\nonumber
I_5 = \langle A_2B_2\rangle - \langle A_2B_0\rangle + \langle A_1B_2\rangle + \langle A_1B_0\rangle \overset{\Omega_5}{\leq} 2,
\end{eqnarray}
\begin{eqnarray}
\nonumber
I_6 = \langle A_2B_1\rangle - \langle A_2B_0\rangle + \langle A_0B_1\rangle + \langle A_0B_0\rangle \overset{\Omega_6}{\leq} 2,
\end{eqnarray}
where $\langle A_xB_y\rangle= \sum_{a,b=0,1} (-1)^{a+b}p(a,b\vert x,y)$ is the expectation value of the measurement outcomes of $a$ and $b$ Alice and Bob given the inputs $x$ and $y$, respectively (similarly for the marginal expectation values $\langle A_x\rangle$ and $\langle B_y\rangle$).

Using the linear program described in the Appendix, we can see how these LF inequalities change if we allow for relaxations of the AOE - our condition \eqref{eq: AOE relaxed}. It follows that the LF bound $\Omega_i$ for each inequality $I_i$ is modified to $\Omega_i^{\epsilon}$ as 
\begin{eqnarray}
\nonumber
\Omega_1^{\epsilon} = 6 + 8\epsilon,
\end{eqnarray}
\begin{eqnarray}
\nonumber
\Omega_2^{\epsilon} = 5 + 8\epsilon,
\end{eqnarray}
\begin{eqnarray}
\nonumber
\Omega_3^{\epsilon} = \Omega_4^{\epsilon} = 4 + 8\epsilon,
\end{eqnarray}
\begin{eqnarray}
\nonumber
\Omega_5^{\epsilon} = \Omega_6^{\epsilon} = 2 + 4\epsilon.
\end{eqnarray}
It follows immediately that a probability distribution $p(a,b\vert x,y)$ achieving a value of $I_i$ given by $\Omega_i^{\epsilon}$ has $\mathcal{A}_c=2\epsilon$. As we argue next, this is the same value for $\mathcal{A}_f$, thus implying that $\mathcal{A}_f=\mathcal{A}_c$ in those cases. To see that, notice that if $I_i=\Omega_i^{\epsilon}$ then the lower bound given by eq. \eqref{eq:af_bound} implies that $\mathcal{A}_f \geq 2 \epsilon$. Further we notice that any decomposition $p(a,b \vert x,y)=q\cdot p_{LF}(a,b \vert x,y) +(1-q) \cdot p_{NS}(a,b \vert x,y)$ provides an upper bound given by $\mathcal{A}_f \leq (1-q)$. In particular, if we choose $p_{NS}$ as a no-signalling (NS) distribution achieving the maximum $\Omega^{\epsilon=1/2}_i$ of inequality $I_i$, $p_{LF}$ as a LF distribution achieving $I_i=\Omega_i$, and make $q=1-2\epsilon$, we obtain that $p(a,b \vert x,y)$ leads to $I_i=\Omega^{\epsilon}_{i}$ with $1-q=2\epsilon$ and thus $\mathcal{A}_f \leq 2\epsilon$. As the lower and upper bounds for $\mathcal{A}_f$ coincide, it follows that $\mathcal{A}_c=\mathcal{A}_f=2\epsilon$.

There are cases, however, where this equivalence between $\mathcal{A}_c$ and $\mathcal{A}_f$ does not hold any longer. To show that, we consider a tripartite scenario. As described in the Appendix, the following symmetry of Mermin's inequality \cite{Mermin1990} is also a valid LF inequality:
\begin{multline}
\nonumber
M = \langle A_3B_3C_2\rangle + \langle A_1B_1C_2\rangle \\+ \langle A_1B_3C_1\rangle - \langle A_3B_1C_1\rangle \leq \beta_{M}= 2.
\end{multline}
If we allow for a relaxation $\epsilon$ of the AOE assumption, it follows that the LF bound is changed as $\Omega_{M}^{\epsilon}=2+8\epsilon$. This implies that a probability distribution achieving $M=\Omega_{M}^{\epsilon}$ requires $\mathcal{A}_c=2\epsilon$. In turn, using the same approach described above for the bipartite case, we can obtain lower and upper bounds for $\mathcal{A}_f$ that coincide and lead to $\mathcal{A}_f=4\epsilon$. That is, in this case, $\mathcal{A}_c=\mathcal{A}_f/2$.

\section{Quantum events are maximally non-absolute}
\label{sec:sec4}
Our main objective is to prove that quantum correlations can be maximally non-absolute according to both measures introduced above, that is, that $\mathcal{A}_f(p_{Q})=1=\mathcal{A}_c(p_{Q})$ for some non-local joint probability distribution $p_{Q}(a,b|x,y)$ in the quantum set \cite{BrunnerEtAl14}. To do so, we first prove a general inequality for the EWFS with dichotomic outcomes but arbitrary measurements and an AOE relaxation as in eqs.\eqref{eq: AOE relaxed}.

\begin{result}
For an EWFS with $k=2$ (2 outcomes), arbitrary $m$ (measurement inputs), and with the AOE relaxed as in eqs.\eqref{eq: AOE relaxed} the following inequality
\begin{eqnarray}
\label{eq: chained}
C^{(m-1)} & := & \langle A_{m-1}B_{m-1}\rangle - \langle A_0B_{m-1}\rangle \\
\nonumber
& + & \sum_{l=0}^{m-2}\left(\langle A_lB_l\rangle + \langle A_{l+1}B_l\rangle\right)\\
\nonumber
& \leq & 2(m - 1) +4\epsilon
\end{eqnarray}
tightly bounds the set $S^{RLF}_{\epsilon}$.
\end{result}

Before proceeding to our main result, it is worthy mentioning that if $\epsilon=0$, that is, in the usual EWFS, the inequality \eqref{eq: chained} is one of the symmetries of the so-called chained Bell inequalities \cite{braunstein1990wringing}.

Next we state our main result.
\begin{result}
Quantum mechanics predicts probabilities distributions $p_{Q}(a,b \vert x,y)$ in the extended Wigner's friend scenario that are maximally non-absolute according to the non-absoluteness coefficient as well as the non-absoluteness fraction. That is, for which $\mathcal{A}_f(p_{Q})=1$ and $\mathcal{A}_c(p_{Q})=1$.
\end{result}
\begin{proof}
The maximal value of $C^{(m-1)}$ allowed in quantum mechanics is \cite{Wehner2006}
\begin{eqnarray}
\nonumber
C^{(m-1)}_{Q_{\mbox{max}}} = 2m\cos\left(\frac{\pi}{2m}\right),
\end{eqnarray}
which can be reached by using a two-qubit system in the entangled state $|\Psi\rangle$,
\begin{eqnarray}
\nonumber
|\Psi\rangle = \frac{1}{\sqrt{2}}\left(|00\rangle + |11\rangle\right),
\end{eqnarray}
shared between Alice and Bob. Alice measures one of the operators $A_j = r_j\sigma_x + s_j\sigma_z$ and Bob $B_j = r_j'\sigma_x + s_j'\sigma_z$ for $j\in\{0,\dots,m - 1\}$. Here, $\sigma_x$ and $\sigma_z$ standing for the Pauli matrices, and
\begin{eqnarray}
\nonumber
r_j & = & \sin\left(\frac{j\pi}{m}\right),\;\;\; s_j = \cos\left(\frac{j\pi}{m}\right),
\end{eqnarray}
and
\begin{eqnarray}
\nonumber
r_j' & = & \sin\left(\frac{(2j + 1)\pi}{m}\right),\;\;\; s_j' = \cos\left(\frac{(2j + 1)\pi}{m}\right).
\end{eqnarray}

Based on the fact that quantum correlations allow for maximal violations of the chained inequality \eqref{eq: chained}, we will first prove that $\mathcal{A}_f=1$. Using the lower bound in eq. \eqref{eq:af_bound} it follows that if   $\Omega_{Q} = \Omega_{NS}$ then $\mathcal{A}_f=1$ and that it is exactly the case of the chained inequality \eqref{eq: chained}, since the maximum quantum violation is equal to the algebraic maximum (and thus also to $\beta_{NS}$) in the limit $m\rightarrow \infty$. This proves that quantum correlations can be maximally non-absolute according to the measure $\mathcal{A}_f$.

The proof that $\mathcal{A}_c=1$ also follows a simple argument. Notice the maximal quantum violation of the chained inequality reaches $C^{m-1}=2m$ a result that requires a relaxation $\epsilon=1/2$ in \eqref{eq: chained} and thus implies maximal non-absoluteness of quantum correlations also regarding the measure $\mathcal{A}_c$.
\end{proof}

\section{Discussion}
\label{sec:sec5}

The interest in the quantum measurement problem and its formulation via Wigner's thought experiment have been recently reignited by a number of results and no-go theorems \cite{brukner2018no,Cavalcanti2021,frauchiger2018quantum,guerin2021no,healey2018quantum,proietti2019experimental,zukowski2021physics,cavalcanti2021view,Bong2020,xu2021nogo}. Similarly to what Bell's theorem \cite{bell1964einstein} did for the EPR experiment \cite{einstein1935can}, these results establish testable constraints of the assumptions -- absoluteness of observed events, no-superdeterminism and parameter independence -- underlying such Wigner's scenarios. Holding on the assumptions of no-superdeterminism and parameter independence, the violation of some constraints, called local friendliness inequalities, are thus experimental witnesses of the non-absoluteness of events. Likewise to the Bell case \cite{chaves2015unifying,de2014nonlocality,brito2018quantifying,wolfe2020quantifying,Brask_2017}, it is natural then to quantify the degree of non-absoluteness of events implied by the violation of a LF inequality.

With that aim in mind, we have introduced two different ways to quantify non-absoluteness. The first measure, called the non-absoluteness fraction $\mathcal{A}_f$, is based on the EPR2 decomposition \cite{elitzur1992quantum} and measures the fraction of measurement outcomes of given experiment that cannot be considered absolute. The second measure, called non-absoluteness coefficient $\mathcal{A}_c$, considers an explicit relaxation of the assumption of absoluteness of observed events and thus allows the predictions of two observers about a given measurement outcome to differ. We have established that $\mathcal{A}_f \leq \mathcal{A}_c$, and have also detailed particular cases where the measures coincide and others where they do not coincide, which shows that both quantities are not trivial.

Following that, we show that a given symmetry \footnote{Any Bell inequality has a number of symmetries, other valid inequalities obtained via the permutation of measurement outcomes, inputs or even the exchange of parties. Those operations, however, are not generally valid when we consider an extended Wigner friend scenario, as now the parties are not on par with each other. For instance, the absoluteness of events constraint specifies "special" inputs, for which the corresponding measurement outcome should be the same as that obtained by the related friend. This would break the symmetry of the inequality related to the permutation of inputs. It is in this sense that we say that even though a given Bell inequality can also be a LF inequality, not all symmetries of that same inequality will have the same property.} of the chained Bell inequality \cite{braunstein1990wringing} is still a valid LF inequality, and also generalized it to the case where relaxations of AOE are allowed. With that, we were able to conclude that quantum correlations can be maximally non-absolute according to the two considered measures, that is, $\mathcal{A}_f=\mathcal{A}_c=1$.

\begin{figure}
    \centering
    \includegraphics[scale = 0.5]{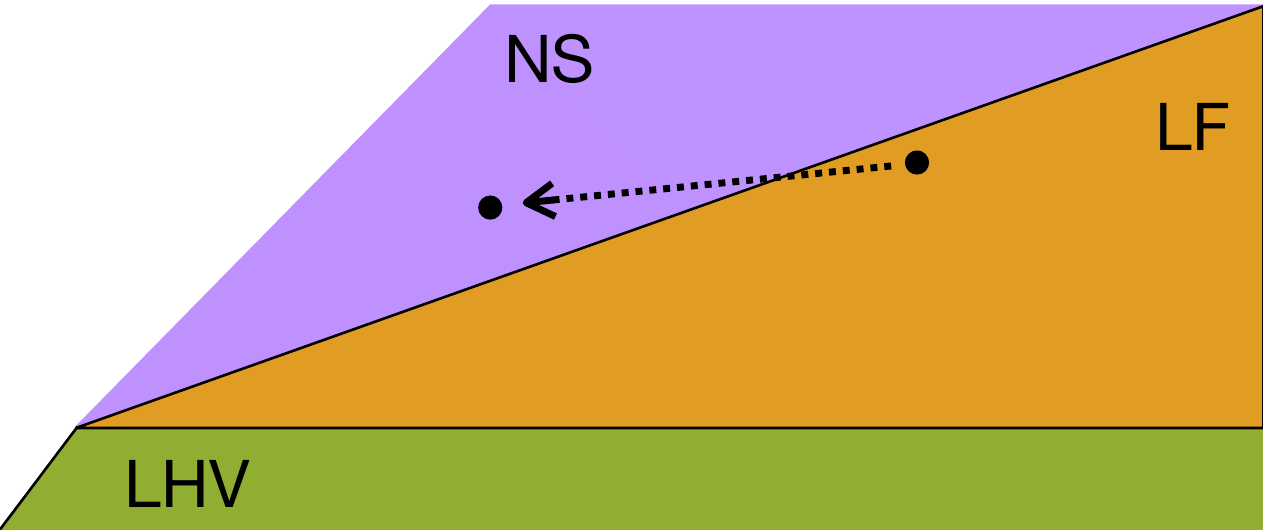}
    \caption{\textbf{Pictorial representation of the set of correlations such that $\mathcal{S}^{LV} \subset \mathcal{S}^{RLF} \subset \mathcal{S}^{NS} $.} Notice that if the set of free operations for non-locality and non-absoluteness differ, it might be possible to use free operations for non-locality that nonetheless might activate the violation of a LF inequality.}
    \label{fig: Concept}
\end{figure}

Notice that any local friendliness inequality is also valid Bell inequality. Theoretically, that is no issue, since the assumptions and the physical scenario is quite different in both cases. This is made clear by the relaxation of AOE that we consider here, a relaxation that simply would make no sense in a Bell scenario. Experimentally, however, the issue is trickier. From a device-independent perspective, the finer details of the experiment and the fact the systems that Alice and Bob measure are actually themselves observers are irrelevant, since Charlie, Debbie, their labs and quantum systems should be treated as a black-box and are thus indistinguishable from a simple common source in a Bell scenario. How can one then decide whether the experimental violation of an LF inequality is not just witnessing the standard Bell nonlocality rather than non-absoluteness of events? One option is to move to a semi-device-independent description where we can distinguish the observers inside the box from a simple source of correlations. Another option, would be to find experimentally testable differences between Bell and the extended Wigner scenarios. For instance, could it be that free-operations for Bell nonlocality can nonetheless increase the level of non-absoluteness of a given observed correlation (see Fig \ref{fig: Concept})? If that would be the case, then one could argue that an experiment where such operations are realized and an increase in the violation of an LF inequality is observed, would be a clear signature for non-absoluteness instead of the good old Bell nonlocality. For that, a resource theory of local friendliness will be needed and we hope our work might motivate attempts in this direction.

\section{Acknowledgements} 
We thank Noah Chaves Leniger for the help with  the figures. This work was supported by The John Templeton Foundation via the grant Q-CAUSAL No 61084 (the opinions expressed in this publication are those of the author(s) and do not necessarily reflect the views of the John Templeton Foundation), by the Serrapilheira Institute (Grant No. Serra-1708-15763), by the Simons Foundation (Grant Number 884966, AF), the Brazilian National Council for Scientific and Technological Development (CNPq) via the National Institute for Science and Technology on Quantum Information (INCT-IQ) and Grants No. 406574/2018-9 and 307295/2020-6, the Brazilian agencies MCTIC, CAPES and MEC. This research was also supported by the Fetzer Franklin Fund of the John E.\ Fetzer Memorial Trust and by grant number FQXi-RFP-IPW-1905 from the Foundational Questions Institute and Fetzer Franklin Fund, a donor advised fund of Silicon Valley Community Foundation.

\onecolumn\newpage
\appendix

\section{The linear program formulation for the relaxation of absoluteness of Observed Events}

A numerical approach to assess the effects of a relaxation on the AOE condition is possible by writing the problem as a linear program. To see this, first we show that the conditions PI and NSD are equivalent to the following condition,
\begin{eqnarray}
\label{eq: NSD and NS}
\left\{\begin{array}{lll}
    p(a,c,d|x,y) & = & p(a,c,d|x),\;\;\forall\;\;a,c,d,x,y, \\
    p(b,c,d|x,y) & = & p(b,c,d|y),\;\;\forall\;\;b,c,d,x,y.
\end{array}\right.
\end{eqnarray}

Indeed, \eqref{eq: NSD} and \eqref{eq: NS} imply \eqref{eq: NSD and NS},
\begin{eqnarray}
\nonumber
p(a|c,d,x,y) & = & p(a|c,d,x)\\
\nonumber
\implies p(a|c,d,x,y)p(c,d|x,y) & = & p(a|c,d,x)p(c,d|x,y)\\
\nonumber
\implies p(a|c,d,x,y)p(c,d|x,y) & = & p(a|c,d,x)p(c,d|x)\\
\nonumber
\implies p(a,c,d|x,y) & = & p(a,c,d|x).
\end{eqnarray}
Also, \eqref{eq: NSD and NS} implies  \eqref{eq: NSD},
\begin{eqnarray}
\label{eq: NDS&NS implies NSD I}
p(a,c,d|x,y) = p(a,c,d|x) \implies p(c,d|x,y) = p(c,d|x),
\end{eqnarray}
and
\begin{eqnarray}
\label{eq: NDS&NS implies NSD II}
p(b,c,d|x,y) = p(b,c,d|y) \implies p(c,d|x,y) = p(c,d|y).
\end{eqnarray}
Finally, \eqref{eq: NSD and NS} implies  \eqref{eq: NS},
\begin{eqnarray}
\nonumber
p(a,c,d|x,y) & = & p(a,c,d|x)\\
\nonumber
\implies p(a|c,d,x,y)p(c,d|x,y) & = & p(a|c,d,x)p(c,d|x)\\
\nonumber
\implies p(a|c,d,x,y)p(c,d|x) & = & p(a|c,d,x)p(c,d|x)\\
\nonumber
\implies p(a|c,d,x,y) & = & p(a|c,d,x).
\end{eqnarray}
Notice that the same approach works by replacing $a$ by $b$.

Using the relation \ref{eq: AOE relaxed}, the problem of finding the upper bound for $I = \sum_{a,b,x,y}\omega_{a,b,x,y}p(a,b|x,y)$ in the presence of a relaxation $\epsilon$, i.e., by a behavior featuring $\mathcal{A}_{c} = 2\epsilon$, can be formally expressed as a linear program:
\begin{align}
    \begin{split}
	\underset{\mathbf{q}\in\RR^{k^{4}*m^{2}}}{\max} \quad & \langle\mathbf{I},\Pi_{a,b,x,y} q\rangle \\
	\text{s.t.} \quad   & \mathbf{q} \geq 0, \\	 
	                    & N\mathbf{q} = \boldsymbol{1}^{(m^2)},\\
	                    & (\Pi_{a,c,d,x,y} - \Pi_{a,c,d,x,y'})\mathbf{q} = \boldsymbol{0}^{(k^2m)}\;\;\;\;\forall a,c,d,x,y,y',\\ 
	                    & (\Pi_{b,c,d,x,y} - \Pi_{b,c,d,x',y})\mathbf{q} = \boldsymbol{0}^{(k^2m)}\;\;\;\;\forall a,c,d,x,x',y,\\
	                    & \Pi_{a=c,x=0,y} \geq (1 - \epsilon)\boldsymbol{1}^{(m)},\\
	                    & \Pi_{b=d,x,y=0} \geq (1 - \epsilon)\boldsymbol{1}^{(m)},
	\label{Optimization}
	\end{split}
\end{align}
in which $\boldsymbol{1}^{(\cdot)}$ is a column vector of the dimension indicated inside the brackets, the entries of which are always $1$ and the vector $\boldsymbol{0}^{(\cdot)}$ is defined analogously. The matrix $N$ is given by,
\begin{eqnarray}
N = \left[\begin{array}{cccc}
    \boldsymbol{1}^{(k^{4}*m^{2})} & \boldsymbol{0}^{(k^{4}*m^{2})} & \boldsymbol{0}^{(k^{4}*m^{2})} & \boldsymbol{0}^{(k^{4}*m^{2})}\\
    \boldsymbol{0}^{(k^{4}*m^{2})} & \boldsymbol{1}^{(k^{4}*m^{2})} & \boldsymbol{0}^{(k^{4}*m^{2})} & \boldsymbol{0}^{(k^{4}*m^{2})}\\
    \boldsymbol{0}^{(k^{4}*m^{2})} & \boldsymbol{0}^{(k^{4}*m^{2})} & \boldsymbol{1}^{(k^{4}*m^{2})} & \boldsymbol{0}^{(k^{4}*m^{2})}\\
    \boldsymbol{0}^{(k^{4}*m^{2})} & \boldsymbol{0}^{(k^{4}*m^{2})} & \boldsymbol{0}^{(k^{4}*m^{2})} & \boldsymbol{1}^{(k^{4}*m^{2})}\\
\end{array}\right]^T.
\end{eqnarray}
The matrix $\Pi_{a,b,x,y}$ projects the vector associated with a distribution $p(a,b,c,d|x,y)$ into the vector associated with its marginal distribution $p(a,b|x,y)$. The matrices $\Pi_{a,c,d,x,y=0}$, $\Pi_{a,c,d,x,y=1}$, $\Pi_{b,c,d,x=0,y}$, and $\Pi_{b,c,d,x=1,y}$ are defined in the same way, while $\Pi_{a=c,x=0,y}$ ($\Pi_{b=d,x,y=0}$) returns a vector containing the probability that $a = c$ ($b = d$) for different values of $y$ ($x$) as entries.

Finally, in equation \eqref{Optimization}, $\mathbf{I}$ is a vector such that $\langle\mathbf{I}, \mathbf{p}\rangle = \sum_{a,b,x,y}\omega_{a,b,x,y}p(a,b|x,y)$, for $\mathbf{p}$ being the vector associated with the distribution $p(a,b|x,y)$.

This analysis can be directly extended into the n-partite case, by generalizing relaxation \eqref{eq: AOE relaxed}, and conditions \eqref{eq: NS} and \eqref{eq: NSD} accordingly. By performing this analysis in the tripartite case, we observe that some symmetries of the Mermin inequality are also valid LF inequalities. For instance given a relaxation of AOE by an amount $\epsilon$ we have that,
\begin{eqnarray}
M = \langle A_3B_3C_2\rangle + \langle A_1B_1C_2\rangle + \langle A_1B_3C_1\rangle - \langle A_3B_1C_1\rangle \leq \Omega_{M}^{\epsilon}=2+8\epsilon.
\end{eqnarray}
in which, $\Omega_{M}^{\epsilon}=2+8\epsilon$.

\section{Proof of Result2}

\begin{proof}
We begin noticing that our $\epsilon-$AOE relaxation is analogous to allowing a certain degree of indeterminism in the scenario described in ref. \cite{Hall2010}.

Consider relaxation \ref{eq: AOE relaxed}, and define the quantities $\alpha_{c,d}$, $\beta_{c,d}$, and $\gamma_{c,d}$,
\begin{eqnarray}
\nonumber
\gamma^{(c,d)}_{x,y} & := & p(a=0,b=0|c,d,x,y),\\
\nonumber
\alpha^{(c,d)}_{x,y} & := & p(a=0|c,d,x,y),\\
\nonumber
\beta^{(c,d)}_{x,y} & := & p(b=0|c,d,x,y).
\label{eq: param_def}
\end{eqnarray}

Positivity implies that
\begin{align}
\label{eq: positivity}
\max \{0, \alpha^{(c,d)}_{x,y} + \beta^{(c,d)}_{x,y} - 1\} \leq \gamma^{(c,d)}_{x,y} \leq \min\{\alpha^{(c,d)}_{x,y},\beta^{(c,d)}_{x,y}\}.
\end{align}

Now define the \emph{conditional correlators} $\langle A_xB_y\rangle_{c,d} = \sum_{a,b} (-1)^{a + b}p(a,b|c,d,x,y)$. It holds that,
\begin{eqnarray}
\nonumber
\langle A_xB_y\rangle_{c,d} & = & 1 + 4\gamma^{(c,d)}_{x,y} - 2\alpha^{(c,d)}_{x,y} - 2\beta^{(c,d)}_{x,y},
\end{eqnarray}
which, combined with inequalities in eq. \eqref{eq: positivity}, and using the fact that $2\max\{r,s\} = r + s + |r - s|$ for any real $r$ and $s$, leads to
\begin{align}
\nonumber
2\left|\alpha^{(c,d)}_{x,y} + \beta^{(c,d)}_{x,y} - 1\right| - 1  \leq \langle A_xB_y\rangle_{c,d} \leq 1 - 2\left|\alpha^{(c,d)}_{x,y} - \beta^{(c,d)}_{x,y}\right|.
\end{align}

Hence the quantity $C_{c,d}^{m-1} = \langle A_{m-1}B_{m-1}\rangle_{c,d} - \langle A_0B_{m-1}\rangle_{c,d} + \sum_{l=0}^{m-2}\left(\langle A_lB_l\rangle_{c,d} + \langle A_{l+1}B_l\rangle_{c,d}\right)$ is bounded,
\begin{eqnarray}
\nonumber
C_{c,d}^{m-1} & \leq & 2 - 2\left|\alpha^{(c,d)}_{m - 1, m - 1} - \beta^{(c,d)}_{m - 1, m - 1}\right| -  2\left|\alpha^{(c,d)}_{0, m - 1} + \beta^{(c,d)}_{0, m - 1} - 1\right|  + \sum_{l=0}^{m - 2} \left(2 - 2\left|\alpha^{(c,d)}_{l,l} - \beta^{(c,d)}_{l,l}\right|\right.\\
\nonumber
& - & \left. 2\left|\alpha^{(c,d)}_{l + 1, l} - \beta^{(c,d)}_{l + 1, l}\right|\right).
\end{eqnarray}

Using condition \ref{eq: NS}, the above expression can be simplified replacing $\alpha^{(c,d)}_{x,y}$ ($p(a=0|c,d,x,y)$) and $\beta^{(c,d)}_{x,y}$ ($p(b=0|c,d,x,y)$), by $\alpha^{(c,d)}_{x}$ ($p(a=0|c,d,x)$) and $\beta^{(c,d)}_{y}$ ($p(b=0|c,d,y)$) respectively,
\begin{eqnarray}
\nonumber
C_{c,d}^{m-1} & \leq & 2m - 2J_{c,d},
\end{eqnarray}
in which
\begin{eqnarray}
\nonumber
J_{c,d} & = & \left|\alpha^{(c,d)}_{m - 1} - \beta^{(c,d)}_{m - 1}\right| + \left|\alpha^{(c,d)}_{0} + \beta^{(c,d)}_{m - 1} - 1\right| + \sum_{l=0}^{m - 2} \left(\left|\alpha^{(c,d)}_{l} - \beta^{(c,d)}_{l}| + |\alpha^{(c,d)}_{l + 1} - \beta^{(c,d)}_{l}\right|\right).
\end{eqnarray}

Notice that:
\begin{align}
\nonumber
\left|\alpha_l^{(c,d)} -\beta_l^{(c,d)}\right| + \left|\alpha_{l+1}^{(c,d)} - \beta_l^{(c,d)}\right| \geq \left|\alpha_l^{(c,d)}-\alpha_{l+1}^{(c,d)}\right|,
\end{align}
with the equality holding true for $\beta_{l}^{(c,d)}$ verifying
\begin{eqnarray}
\label{eq: beta opt}
\beta_l^{(c,d)}= \frac{1}{2}\left(\alpha_l^{(c,d)}+\alpha_{l+1}^{(c,d)}\right) \;\;\mbox{for } l \leq m-2.
\end{eqnarray}
Which leads to:
\begin{eqnarray}
\nonumber
J_{c,d} & \geq & \left|\alpha^{(c,d)}_{m - 1} - \beta^{(c,d)}_{m - 1}\right| + \left|\alpha^{(c,d)}_{0} + \beta^{(c,d)}_{m - 1} - 1\right| + \sum_{l=0}^{m - 2} \left(\left|\alpha^{(c,d)}_{l} - \alpha^{(c,d)}_{l + 1} \right|\right), 
\label{eq: Jbound_beta}
\end{eqnarray}
equality being reached via equation \eqref{eq: beta opt}.

On the other hand, we can perform a similar analysis considering $\alpha_l^{(c,d)}$ for $l\in\{1,\dots,m-2\}$. 
Similarly to the previous case, we can use the triangle inequality successively in the terms within the summation in Eq.\ \eqref{eq: Jbound_beta} to conclude that
\begin{equation}
\sum_{l=0}^{m-2} \left| \alpha_l^{(c,d)} - \alpha_{l+1}^{(c,d)} \right| \geq \left|\alpha_0^{(c,d)} - \alpha_{m-1}^{(c,d)}\right|.    
\end{equation}
We may saturate this lower bound by setting $\alpha_l^{(c,d)}$ as
\begin{equation}
\label{eq: alpha opt}
\alpha_l^{(c,d)} = \frac{l}{m-1} \alpha_{m-1}^{(c,d)} + \frac{m-1-l}{m-1} \alpha_0^{(c,d)},
\end{equation}
for $l \in \{1,\ldots,m-2\}$, which can easily be verified, for instance by noticing that $\alpha_{l+1}^{(c,d)}  - \alpha_l^{(c,d)} = (\alpha_{m-1}^{(c,d)} - \alpha_0^{(c,d)})/(m-1)$. It should be noted that this choice of $\alpha_l^{(c,d)}$ is equivalent to $\alpha_l^{(c,d)} = (\alpha_{l-1}^{(c,d)} + \alpha_{l+1}^{(c,d)})/2$, which follows from the same reasoning as used for $\beta_l^{(c,d)}$.

Using the triangle inequality again, we can find the optimal value of $\alpha^{(c,d)}_{0}$:
\begin{align}
\left|\alpha^{(c,d)}_{0} - \left(1-\beta^{(c,d)}_{m - 1}\right)\right| + \left|\alpha^{(c,d)}_{m - 1} - \alpha^{(c,d)}_{0}\right| \geq   \left|\beta^{(c,d)}_{m - 1} - 1 + \alpha^{(c,d)}_{m - 1}\right|,
\end{align}
with the equality holding true when,
\begin{eqnarray}
\label{eq: alpha_0 opt}
\alpha^{(c,d)}_{0} = \frac{1}{2}\left(1 - \beta_{m - 1}^{(c,d)} + \alpha^{(c,d)}_{m - 1}\right).
\end{eqnarray}

From which we conclude that,
\begin{eqnarray}
\nonumber
J_{c,d} & \geq & \left|\alpha^{(c,d)}_{m - 1} - \beta^{(c,d)}_{m - 1}\right| + \left|\alpha^{(c,d)}_{m - 1} + \beta^{(c,d)}_{m - 1} - 1 \right|.
\label{eq: J_bound}
\end{eqnarray}

Using the triangle inequality one last time, we can see that
\begin{align}
\nonumber
\left|\alpha^{(c,d)}_{m - 1} - \beta^{(c,d)}_{m - 1}\right| + \left|\alpha^{(c,d)}_{m - 1} + \beta^{(c,d)}_{m - 1} - 1 \right| \geq \left|2\alpha^{(c,d)}_{m - 1} - 1 \right|,
\end{align}
which can be achieved by setting $\beta^{(c,d)}_{m - 1} = \alpha^{(c,d)}_{m - 1}$, leading to the following tight relation:
\begin{eqnarray}
\nonumber
J_{c,d} & \geq & \left|2\alpha^{(c,d)}_{m - 1} - 1 \right|
\end{eqnarray}

We highlight that the values for which the above equation becomes an equality are always well-defined. From equation \eqref{eq: alpha_0 opt}, one gets that $\alpha^{(c,d)}_{0}\in[0,1]$ as long as $0\leq\alpha_{m - 1}^{(c,d)}\leq1$ and $0\leq 1 - \beta_0^{(c,d)}\leq 1$ -- which is always the case. From equation \eqref{eq: alpha opt}, for $1\leq l\leq m-2$, $\alpha_l^{(c,d)}\in[0,1]$ as long as $0\leq\alpha_{m - 1}^{(c,d)}\leq1$ and $0 \leq\alpha^{(c,d)}_{0}\leq 1$ -- which is always the case. The same holds for $\beta_l^{(c,d)}$ in the case $l>0$.

The above result leads to the following condition
\begin{eqnarray}
C^{m-1} = \sum_{c,d}p(c,d)C^{m-1}_{c,d}\leq 2m - 2J,
\end{eqnarray}
$J$ being a convex combination of $J_{c,d}$,
\begin{eqnarray}
\nonumber
J & = & \sum_{c,d}p(c,d)J_{c,d}\\
& \geq &\sum_{c,d}p(c,d)\left|2\alpha^{(c,d)}_{m - 1} - 1 \right|.
\end{eqnarray}
This means that the minimum value of $J$ is obtained by setting $p(c^*,d^*)=1$, in which $c^*$ and $d^*$ are such that $\left|2\alpha^{(c^*,d^*)}_{m - 1} - 1 \right|\geq \left|2\alpha^{(c,d)}_{m - 1} - 1 \right|$, for all $c$ and $d$. Which implies that\begin{eqnarray}
\nonumber
J & \geq & \left|2p(a=0,c^*,d^*|x=m-1) - p(c^*,d^*) \right|\\
\nonumber
& = & \left|2p(a=0,c^*,d^*|x=m-1) - 1\right|\\
\nonumber
& = & \left|2p(a=0,c^*|x=m-1) - 1\right|\\
& = & \left|2p(a=0,c^*|x=m-1,y) - 1\right|,
\end{eqnarray}
in which we have set $p(c^*,d^*) = 1$.

If $c^*=0$, then $p(a=0,c^*|x=m-1) \in [0, 1 - \epsilon]$ and $J\geq 1- 2\epsilon$, on the other hand, the case in which $c^*=1$ imply that $p(a=0,c^*|x=m-1) \in [0,\epsilon]$, which also lead to $J\geq 1- 2\epsilon$. Thus concluding that the following tight bound holds 
\begin{eqnarray}
C^{m-1}\leq 2(m-1) + 4\epsilon.
\end{eqnarray}
\end{proof}

\section{Alternative proof that the Chained Bell inequality is also a valid LF inequality}

\begin{result}
The EWFS with $k=2$ (2 outcomes) and arbitrary $m$ (measurement inputs) is bounded by the following LF inequality:
\begin{eqnarray}
C^{(m-1)}_j = \langle A_{m-1}B_{m-1}\rangle - \langle A_jB_{m-1}\rangle + \sum_{l=j}^{m-2}\left(\langle A_lB_l\rangle + \langle A_{l+1}B_l\rangle\right)\leq 2((m-j)-1)
\end{eqnarray}
for j=0.
\end{result}
\begin{proof}
Before proceeding with the proof, it is worthy mentioning that for each $j$ and $m$ the inequality above is one of the symmetries of the so-called chained Bell inequalities \cite{braunstein1990wringing}. That said, we write this inequality in a slightly different way of how it is typically presented (see for instance \cite{Supic2016}), which can be recovered from our expression by setting $j=0$ and, by defining $\langle A_{m}B_{m-1}\rangle = - \langle A_{0}B_{m-1}\rangle$.

Now consider the following symmetries of the chained inequality, which bound the Bell scenario for $k=2$ and arbitrary $m$,
\begin{eqnarray}
C^{(m-1)}_j = \langle A_{m-1}B_{m-1}\rangle - \langle A_jB_{m-1}\rangle + \sum_{l=j}^{m-2}\left(\langle A_lB_l\rangle + \langle A_{l+1}B_l\rangle\right)\leq 2((m-j)-1)
\end{eqnarray}
Putting the term with $l=j$ outside the summation we obtain
\begin{eqnarray}
\nonumber
C^{(m - 1)}_j & = & \langle A_{m - 1}B_{m - 1}\rangle - \langle A_jB_{m - 1}\rangle + \langle A_jB_j\rangle + \langle A_{j + 1}B_j\rangle + \sum_{l=j + 1}^{m - 2}\left(\langle A_lB_l\rangle + \langle A_{l + 1}B_l\rangle\right),
\end{eqnarray}
and adding $0 = \langle A_{j+1}B_{m - 1}\rangle -\langle A_{j + 1}B_{m - 1}\rangle$ to it we get
\begin{eqnarray}
\nonumber
C^{(m - 1)}_j & = & \langle A_{j + 1}B_{m - 1}\rangle -\langle A_{j + 1}B_{m - 1}\rangle + \langle A_{m - 1}B_{m - 1}\rangle - \langle A_jB_{m - 1}\rangle + \langle A_jB_j\rangle + \langle A_{j + 1}B_j\rangle \\
\nonumber
& + & \sum_{l = j + 1}^{m - 2}\left(\langle A_lB_l\rangle + \langle A_{l + 1}B_l\rangle\right).
\end{eqnarray}
Now using the very definition of $C^{(m-1)}_j$ we obtain
\begin{eqnarray}
\nonumber
C^{(m - 1)}_j & = & \langle A_{j + 1}B_{m - 1}\rangle - \langle A_jB_{m - 1}\rangle + \langle A_jB_j\rangle + \langle A_{j + 1}B_j\rangle + C^{(m - 1)}_{j + 1}\\
\nonumber
& = & \tilde{C}^{(m - 1)}_j + C^{(m - 1)}_{j + 1},
\end{eqnarray}
in which
\begin{eqnarray}
\tilde{C}_j^{(m - 1)} = \langle A_{j + 1}B_{m - 1}\rangle - \langle A_j B_{m - 1}\rangle + \langle A_jB_j\rangle + \langle A_{j + 1}B_j\rangle,
\end{eqnarray}
is also a symmetry of the chained inequality with only two inputs.

Using this recurrence relation we can rewrite
\begin{eqnarray}
\nonumber
C^{(m - 1)}_0 & = & \tilde{C}^{(m - 1)}_0 + C^{(m - 1)}_{1}\\
\nonumber
& = & \tilde{C}^{(m - 1)}_0 + \tilde{C}^{(m - 1)}_1 + C^{(m - 1)}_{2}\\
\nonumber
& \vdots&\\
\nonumber
& = & C^{(m - 1)}_{m - 2} + \sum_{j=0}^{m - 3}\tilde{C}^{m - 1}_{j}.
\end{eqnarray}

By the results of reference \cite{Bong2020}, we know that in the EWFS, certain symmetries of the CHSH inequality are a valid LF inequalities. In particular it holds that
\begin{eqnarray}
\tilde{C}^{(m - 1)}_{j}\leq 2\;\;\;\; \forall \; j,
\end{eqnarray}
and
\begin{eqnarray}
C^{(m - 1)}_{m-2}\leq 2,
\end{eqnarray}
which implies that
\begin{eqnarray}
\nonumber
C^{(m - 1)}_0 & \leq & 2 + \sum_{j=0}^{m - 3}2\\
& = & 2(m - 1),
\end{eqnarray}
concluding the proof.
\end{proof}


\begin{thebibliography}{45}%
\makeatletter
\providecommand \@ifxundefined [1]{%
 \@ifx{#1\undefined}
}%
\providecommand \@ifnum [1]{%
 \ifnum #1\expandafter \@firstoftwo
 \else \expandafter \@secondoftwo
 \fi
}%
\providecommand \@ifx [1]{%
 \ifx #1\expandafter \@firstoftwo
 \else \expandafter \@secondoftwo
 \fi
}%
\providecommand \natexlab [1]{#1}%
\providecommand \enquote  [1]{``#1''}%
\providecommand \bibnamefont  [1]{#1}%
\providecommand \bibfnamefont [1]{#1}%
\providecommand \citenamefont [1]{#1}%
\providecommand \href@noop [0]{\@secondoftwo}%
\providecommand \href [0]{\begingroup \@sanitize@url \@href}%
\providecommand \@href[1]{\@@startlink{#1}\@@href}%
\providecommand \@@href[1]{\endgroup#1\@@endlink}%
\providecommand \@sanitize@url [0]{\catcode `\\12\catcode `\$12\catcode
  `\&12\catcode `\#12\catcode `\^12\catcode `\_12\catcode `\%12\relax}%
\providecommand \@@startlink[1]{}%
\providecommand \@@endlink[0]{}%
\providecommand \url  [0]{\begingroup\@sanitize@url \@url }%
\providecommand \@url [1]{\endgroup\@href {#1}{\urlprefix }}%
\providecommand \urlprefix  [0]{URL }%
\providecommand \Eprint [0]{\href }%
\providecommand \doibase [0]{https://doi.org/}%
\providecommand \selectlanguage [0]{\@gobble}%
\providecommand \bibinfo  [0]{\@secondoftwo}%
\providecommand \bibfield  [0]{\@secondoftwo}%
\providecommand \translation [1]{[#1]}%
\providecommand \BibitemOpen [0]{}%
\providecommand \bibitemStop [0]{}%
\providecommand \bibitemNoStop [0]{.\EOS\space}%
\providecommand \EOS [0]{\spacefactor3000\relax}%
\providecommand \BibitemShut  [1]{\csname bibitem#1\endcsname}%
\let\auto@bib@innerbib\@empty
\bibitem [{\citenamefont {Wigner}(1963)}]{wigner1963problem}%
  \BibitemOpen
  \bibfield  {author} {\bibinfo {author} {\bibfnamefont {E.~P.}\ \bibnamefont
  {Wigner}},\ }\bibfield  {title} {\bibinfo {title} {The problem of
  measurement},\ }\href {https://doi.org/10.1119/1.1969254} {\bibfield  {journal} {\bibinfo  {journal}
  {American Journal of Physics}\ }\textbf {\bibinfo {volume} {31}},\ \bibinfo
  {pages} {6} (\bibinfo {year} {1963})}\BibitemShut {NoStop}%
\bibitem [{\citenamefont {Schlosshauer}(2005)}]{schlosshauer2005decoherence}%
  \BibitemOpen
  \bibfield  {author} {\bibinfo {author} {\bibfnamefont {M.}~\bibnamefont
  {Schlosshauer}},\ }\bibfield  {title} {\bibinfo {title} {Decoherence, the
  measurement problem, and interpretations of quantum mechanics},\ }\href
  {https://doi.org/10.1103/RevModPhys.76.1267} {\bibfield  {journal} {\bibinfo  {journal} {Reviews of Modern physics}\
  }\textbf {\bibinfo {volume} {76}},\ \bibinfo {pages} {1267} (\bibinfo {year}
  {2005})}\BibitemShut {NoStop}%
\bibitem [{\citenamefont {Pusey}(2018)}]{Pusey18}%
  \BibitemOpen
  \bibfield  {author} {\bibinfo {author} {\bibfnamefont {M.~F.}\ \bibnamefont
  {Pusey}},\ }\bibfield  {title} {\bibinfo {title} {An inconsistent friend},\
  }\href {https://doi.org/10.1038/s41567-018-0293-7} {\bibfield  {journal}
  {\bibinfo  {journal} {Nature Physics}\ }\textbf {\bibinfo {volume} {14}},\
  \bibinfo {pages} {977--978} (\bibinfo {year} {2018})}\BibitemShut {NoStop}%
\bibitem [{\citenamefont {Wigner}(1995)}]{wigner1995remarks}%
  \BibitemOpen
  \bibfield  {author} {\bibinfo {author} {\bibfnamefont {E.~P.}\ \bibnamefont
  {Wigner}},\ }\bibfield  {title} {\bibinfo {title} {Remarks on the mind-body
  question},\ }in\ \href {https://doi.org/10.1007/978-3-642-78374-6_20} {\emph {\bibinfo {booktitle} {Philosophical
  reflections and syntheses}}}\ (\bibinfo  {publisher} {Springer},\ \bibinfo
  {year} {1995})\ pp.\ \bibinfo {pages} {247--260}\BibitemShut {NoStop}%
\bibitem [{\citenamefont {Everett}(2015)}]{everett2015relative}%
  \BibitemOpen
  \bibfield  {author} {\bibinfo {author} {\bibfnamefont {H.}~\bibnamefont
  {Everett}},\ }\bibfield  {title} {\bibinfo {title} {"Relative state"
  formulation of quantum mechanics},\ }\href {https://doi.org/10.1515/9781400868056-003} {\bibfield  {journal}
  {\bibinfo  {journal} {The Many Worlds Interpretation of Quantum Mechanics}\
  ,\ \bibinfo {pages} {141}} (\bibinfo {year} {2015})}\BibitemShut {NoStop}%
\bibitem [{\citenamefont {Bohm}\ and\ \citenamefont
  {Bub}(1966)}]{bohm1966proposed}%
  \BibitemOpen
  \bibfield  {author} {\bibinfo {author} {\bibfnamefont {D.}~\bibnamefont
  {Bohm}}\ and\ \bibinfo {author} {\bibfnamefont {J.}~\bibnamefont {Bub}},\
  }\bibfield  {title} {\bibinfo {title} {A proposed solution of the measurement
  problem in quantum mechanics by a hidden variable theory},\ }\href {https://doi.org/10.1103/RevModPhys.38.453}
  {\bibfield  {journal} {\bibinfo  {journal} {Reviews of Modern Physics}\
  }\textbf {\bibinfo {volume} {38}},\ \bibinfo {pages} {453} (\bibinfo {year}
  {1966})}\BibitemShut {NoStop}%
\bibitem [{\citenamefont {Hossenfelder}\ and\ \citenamefont
  {Palmer}(2020)}]{hossenfelder2020rethinking}%
  \BibitemOpen
  \bibfield  {author} {\bibinfo {author} {\bibfnamefont {S.}~\bibnamefont
  {Hossenfelder}}\ and\ \bibinfo {author} {\bibfnamefont {T.}~\bibnamefont
  {Palmer}},\ }\bibfield  {title} {\bibinfo {title} {Rethinking
  superdeterminism},\ }\href {https://doi.org/10.3389/fphy.2020.00139} {\bibfield  {journal} {\bibinfo  {journal}
  {Frontiers in Physics}\ }\textbf {\bibinfo {volume} {8}},\ \bibinfo {pages}
  {139} (\bibinfo {year} {2020})}\BibitemShut {NoStop}%
\bibitem [{\citenamefont {Hooft}(2007)}]{hooft2007free}%
  \BibitemOpen
  \bibfield  {author} {\bibinfo {author} {\bibfnamefont {G.}~\bibnamefont
  {Hooft}},\ }\bibfield  {title} {\bibinfo {title} {The free-will postulate in
  quantum mechanics},\ }\href {https://doi.org/10.48550/arXiv.quant-ph/0701097} {\bibfield  {journal} {\bibinfo
  {journal} {arXiv preprint quant-ph/0701097}\ } (\bibinfo {year}
  {2007})}\BibitemShut {NoStop}%
\bibitem [{\citenamefont {Price}(2008)}]{price2008toy}%
  \BibitemOpen
  \bibfield  {author} {\bibinfo {author} {\bibfnamefont {H.}~\bibnamefont
  {Price}},\ }\bibfield  {title} {\bibinfo {title} {Toy models for
  retrocausality},\ }\href {https://doi.org/10.1016/j.shpsb.2008.05.006} {\bibfield  {journal} {\bibinfo  {journal}
  {Studies in History and Philosophy of Science Part B: Studies in History and
  Philosophy of Modern Physics}\ }\textbf {\bibinfo {volume} {39}},\ \bibinfo
  {pages} {752} (\bibinfo {year} {2008})}\BibitemShut {NoStop}%
\bibitem [{\citenamefont {Stapp}(1972)}]{stapp1972copenhagen}%
  \BibitemOpen
  \bibfield  {author} {\bibinfo {author} {\bibfnamefont {H.~P.}\ \bibnamefont
  {Stapp}},\ }\bibfield  {title} {\bibinfo {title} {The copenhagen
  interpretation},\ }\href {https://doi.org/10.1119/1.1986768} {\bibfield  {journal} {\bibinfo  {journal}
  {American journal of physics}\ }\textbf {\bibinfo {volume} {40}},\ \bibinfo
  {pages} {1098} (\bibinfo {year} {1972})}\BibitemShut {NoStop}%
\bibitem [{\citenamefont {Rovelli}(1996)}]{rovelli1996relational}%
  \BibitemOpen
  \bibfield  {author} {\bibinfo {author} {\bibfnamefont {C.}~\bibnamefont
  {Rovelli}},\ }\bibfield  {title} {\bibinfo {title} {Relational quantum
  mechanics},\ }\href {https://doi.org/10.1007/BF02302261} {\bibfield  {journal} {\bibinfo  {journal}
  {International Journal of Theoretical Physics}\ }\textbf {\bibinfo {volume}
  {35}},\ \bibinfo {pages} {1637} (\bibinfo {year} {1996})}\BibitemShut
  {NoStop}%
\bibitem [{\citenamefont {Caves}\ \emph {et~al.}(2002)\citenamefont {Caves},
  \citenamefont {Fuchs},\ and\ \citenamefont {Schack}}]{caves2002quantum}%
  \BibitemOpen
  \bibfield  {author} {\bibinfo {author} {\bibfnamefont {C.~M.}\ \bibnamefont
  {Caves}}, \bibinfo {author} {\bibfnamefont {C.~A.}\ \bibnamefont {Fuchs}},\
  and\ \bibinfo {author} {\bibfnamefont {R.}~\bibnamefont {Schack}},\
  }\bibfield  {title} {\bibinfo {title} {Quantum probabilities as bayesian
  probabilities},\ }\href {https://doi.org/10.1103/PhysRevA.65.022305} {\bibfield  {journal} {\bibinfo  {journal}
  {Physical review A}\ }\textbf {\bibinfo {volume} {65}},\ \bibinfo {pages}
  {022305} (\bibinfo {year} {2002})}\BibitemShut {NoStop}%
\bibitem [{\citenamefont {Bassi}\ and\ \citenamefont
  {Ghirardi}(2003)}]{bassi2003dynamical}%
  \BibitemOpen
  \bibfield  {author} {\bibinfo {author} {\bibfnamefont {A.}~\bibnamefont
  {Bassi}}\ and\ \bibinfo {author} {\bibfnamefont {G.}~\bibnamefont
  {Ghirardi}},\ }\bibfield  {title} {\bibinfo {title} {Dynamical reduction
  models},\ }\href {https://doi.org/10.1016/S0370-1573(03)00103-0} {\bibfield  {journal} {\bibinfo  {journal} {Physics
  Reports}\ }\textbf {\bibinfo {volume} {379}},\ \bibinfo {pages} {257}
  (\bibinfo {year} {2003})}\BibitemShut {NoStop}%
\bibitem [{\citenamefont {Ghirardi}\ \emph {et~al.}(1986)\citenamefont
  {Ghirardi}, \citenamefont {Rimini},\ and\ \citenamefont
  {Weber}}]{ghirardi1986unified}%
  \BibitemOpen
  \bibfield  {author} {\bibinfo {author} {\bibfnamefont {G.~C.}\ \bibnamefont
  {Ghirardi}}, \bibinfo {author} {\bibfnamefont {A.}~\bibnamefont {Rimini}},\
  and\ \bibinfo {author} {\bibfnamefont {T.}~\bibnamefont {Weber}},\ }\bibfield
  {title} {\bibinfo {title} {Unified dynamics for microscopic and macroscopic
  systems},\ }\href {https://doi.org/10.1103/PhysRevD.34.470} {\bibfield  {journal} {\bibinfo  {journal}
  {Physical review D}\ }\textbf {\bibinfo {volume} {34}},\ \bibinfo {pages}
  {470} (\bibinfo {year} {1986})}\BibitemShut {NoStop}%
\bibitem [{\citenamefont {Penrose}(1996)}]{penrose1996gravity}%
  \BibitemOpen
  \bibfield  {author} {\bibinfo {author} {\bibfnamefont {R.}~\bibnamefont
  {Penrose}},\ }\bibfield  {title} {\bibinfo {title} {On gravity's role in
  quantum state reduction},\ }\href {https://doi.org/10.1007/BF02105068} {\bibfield  {journal} {\bibinfo
  {journal} {General relativity and gravitation}\ }\textbf {\bibinfo {volume}
  {28}},\ \bibinfo {pages} {581} (\bibinfo {year} {1996})}\BibitemShut
  {NoStop}%
\bibitem [{\citenamefont {Brukner}(2015)}]{Brukner2015}%
  \BibitemOpen
  \bibfield  {author} {\bibinfo {author} {\bibfnamefont {C.}~\bibnamefont
  {Brukner}},\ }\href {https://doi.org/10.48550/arXiv.1507.05255} {\bibinfo {title} {On the quantum measurement
  problem}} (\bibinfo {year} {2015}),\  {arXiv:1507.05255 [quant-ph]} \BibitemShut
  {NoStop}%
\bibitem [{\citenamefont {Brukner}(2018)}]{brukner2018no}%
  \BibitemOpen
  \bibfield  {author} {\bibinfo {author} {\bibfnamefont {{\v{C}}.}~\bibnamefont
  {Brukner}},\ }\bibfield  {title} {\bibinfo {title} {A no-go theorem for
  observer-independent facts},\ }\href {https://doi.org/10.3390/e20050350} {\bibfield  {journal} {\bibinfo
  {journal} {Entropy}\ }\textbf {\bibinfo {volume} {20}},\ \bibinfo {pages}
  {350} (\bibinfo {year} {2018})}\BibitemShut {NoStop}%
\bibitem [{\citenamefont {Cavalcanti}\ and\ \citenamefont
  {Wiseman}(2021)}]{Cavalcanti2021}%
  \BibitemOpen
  \bibfield  {author} {\bibinfo {author} {\bibfnamefont {E.~G.}\ \bibnamefont
  {Cavalcanti}}\ and\ \bibinfo {author} {\bibfnamefont {H.~M.}\ \bibnamefont
  {Wiseman}},\ }\bibfield  {title} {\bibinfo {title} {Implications of local
  friendliness violation for quantum causality},\ }\bibfield  {journal}
  {\bibinfo  {journal} {Entropy}\ }\textbf {\bibinfo {volume} {23}},\ \href
  {https://doi.org/10.3390/e23080925} {10.3390/e23080925} (\bibinfo {year}
  {2021})\BibitemShut {NoStop}%
\bibitem [{\citenamefont {Frauchiger}\ and\ \citenamefont
  {Renner}(2018)}]{frauchiger2018quantum}%
  \BibitemOpen
  \bibfield  {author} {\bibinfo {author} {\bibfnamefont {D.}~\bibnamefont
  {Frauchiger}}\ and\ \bibinfo {author} {\bibfnamefont {R.}~\bibnamefont
  {Renner}},\ }\bibfield  {title} {\bibinfo {title} {Quantum theory cannot
  consistently describe the use of itself},\ }\href {https://doi.org/10.1038/s41467-018-05739-8} {\bibfield
  {journal} {\bibinfo  {journal} {Nature communications}\ }\textbf {\bibinfo
  {volume} {9}},\ \bibinfo {pages} {1} (\bibinfo {year} {2018})}\BibitemShut
  {NoStop}%
\bibitem [{\citenamefont {Gu{\'e}rin}\ \emph {et~al.}(2021)\citenamefont
  {Gu{\'e}rin}, \citenamefont {Baumann}, \citenamefont {Del~Santo},\ and\
  \citenamefont {Brukner}}]{guerin2021no}%
  \BibitemOpen
  \bibfield  {author} {\bibinfo {author} {\bibfnamefont {P.~A.}\ \bibnamefont
  {Gu{\'e}rin}}, \bibinfo {author} {\bibfnamefont {V.}~\bibnamefont {Baumann}},
  \bibinfo {author} {\bibfnamefont {F.}~\bibnamefont {Del~Santo}},\ and\
  \bibinfo {author} {\bibfnamefont {{\v{C}}.}~\bibnamefont {Brukner}},\
  }\bibfield  {title} {\bibinfo {title} {A no-go theorem for the persistent
  reality of Wigner's friends perception},\ }\href {https://doi.org/10.1038/s42005-021-00589-1} {\bibfield
  {journal} {\bibinfo  {journal} {Communications Physics}\ }\textbf {\bibinfo
  {volume} {4}},\ \bibinfo {pages} {1} (\bibinfo {year} {2021})}\BibitemShut
  {NoStop}%
\bibitem [{\citenamefont {Healey}(2018)}]{healey2018quantum}%
  \BibitemOpen
  \bibfield  {author} {\bibinfo {author} {\bibfnamefont {R.}~\bibnamefont
  {Healey}},\ }\bibfield  {title} {\bibinfo {title} {Quantum theory and the
  limits of objectivity},\ }\href {https://doi.org/10.1007/s10701-018-0216-6} {\bibfield  {journal} {\bibinfo
  {journal} {Foundations of Physics}\ }\textbf {\bibinfo {volume} {48}},\
  \bibinfo {pages} {1568} (\bibinfo {year} {2018})}\BibitemShut {NoStop}%
\bibitem [{\citenamefont {Proietti}\ \emph {et~al.}(2019)\citenamefont
  {Proietti}, \citenamefont {Pickston}, \citenamefont {Graffitti},
  \citenamefont {Barrow}, \citenamefont {Kundys}, \citenamefont {Branciard},
  \citenamefont {Ringbauer},\ and\ \citenamefont
  {Fedrizzi}}]{proietti2019experimental}%
  \BibitemOpen
  \bibfield  {author} {\bibinfo {author} {\bibfnamefont {M.}~\bibnamefont
  {Proietti}}, \bibinfo {author} {\bibfnamefont {A.}~\bibnamefont {Pickston}},
  \bibinfo {author} {\bibfnamefont {F.}~\bibnamefont {Graffitti}}, \bibinfo
  {author} {\bibfnamefont {P.}~\bibnamefont {Barrow}}, \bibinfo {author}
  {\bibfnamefont {D.}~\bibnamefont {Kundys}}, \bibinfo {author} {\bibfnamefont
  {C.}~\bibnamefont {Branciard}}, \bibinfo {author} {\bibfnamefont
  {M.}~\bibnamefont {Ringbauer}},\ and\ \bibinfo {author} {\bibfnamefont
  {A.}~\bibnamefont {Fedrizzi}},\ }\bibfield  {title} {\bibinfo {title}
  {Experimental test of local observer independence},\ }\href {https://doi.org/10.1126/sciadv.aaw9832}
  {\bibfield  {journal} {\bibinfo  {journal} {Science advances}\ }\textbf
  {\bibinfo {volume} {5}},\ \bibinfo {pages} {eaaw9832} (\bibinfo {year}
  {2019})}\BibitemShut {NoStop}%
\bibitem [{\citenamefont {{\.Z}ukowski}\ and\ \citenamefont
  {Markiewicz}(2021)}]{zukowski2021physics}%
  \BibitemOpen
  \bibfield  {author} {\bibinfo {author} {\bibfnamefont {M.}~\bibnamefont
  {{\.Z}ukowski}}\ and\ \bibinfo {author} {\bibfnamefont {M.}~\bibnamefont
  {Markiewicz}},\ }\bibfield  {title} {\bibinfo {title} {Physics and
  metaphysics of Wigner's friends: Even performed premeasurements have no
  results},\ }\href {https://doi.org/10.1103/PhysRevLett.126.130402} {\bibfield  {journal} {\bibinfo  {journal}
  {Physical Review Letters}\ }\textbf {\bibinfo {volume} {126}},\ \bibinfo
  {pages} {130402} (\bibinfo {year} {2021})}\BibitemShut {NoStop}%
\bibitem [{\citenamefont {Cavalcanti}(2021)}]{cavalcanti2021view}%
  \BibitemOpen
  \bibfield  {author} {\bibinfo {author} {\bibfnamefont {E.~G.}\ \bibnamefont
  {Cavalcanti}},\ }\bibfield  {title} {\bibinfo {title} {The view from a Wigner
  bubble},\ }\href {https://doi.org/10.1007/s10701-021-00417-0} {\bibfield  {journal} {\bibinfo  {journal}
  {Foundations of Physics}\ }\textbf {\bibinfo {volume} {51}},\ \bibinfo
  {pages} {1} (\bibinfo {year} {2021})}\BibitemShut {NoStop}%
\bibitem [{\citenamefont {Bong}\ \emph {et~al.}(2020)\citenamefont {Bong},
  \citenamefont {Utreras-Alarc{\'o}n}, \citenamefont {Ghafari}, \citenamefont
  {Liang}, \citenamefont {Tischler}, \citenamefont {Cavalcanti}, \citenamefont
  {Pryde},\ and\ \citenamefont {Wiseman}}]{Bong2020}%
  \BibitemOpen
  \bibfield  {author} {\bibinfo {author} {\bibfnamefont {K.-W.}\ \bibnamefont
  {Bong}}, \bibinfo {author} {\bibfnamefont {A.}~\bibnamefont
  {Utreras-Alarc{\'o}n}}, \bibinfo {author} {\bibfnamefont {F.}~\bibnamefont
  {Ghafari}}, \bibinfo {author} {\bibfnamefont {Y.-C.}\ \bibnamefont {Liang}},
  \bibinfo {author} {\bibfnamefont {N.}~\bibnamefont {Tischler}}, \bibinfo
  {author} {\bibfnamefont {E.~G.}\ \bibnamefont {Cavalcanti}}, \bibinfo
  {author} {\bibfnamefont {G.~J.}\ \bibnamefont {Pryde}},\ and\ \bibinfo
  {author} {\bibfnamefont {H.~M.}\ \bibnamefont {Wiseman}},\ }\bibfield
  {title} {\bibinfo {title} {A strong no-go theorem on the Wigner's friend
  paradox},\ }\href {https://doi.org/10.1038/s41567-020-0990-x} {\bibfield
  {journal} {\bibinfo  {journal} {Nature Physics}\ }\textbf {\bibinfo {volume}
  {16}},\ \bibinfo {pages} {1199} (\bibinfo {year} {2020})}\BibitemShut
  {NoStop}%
\bibitem [{\citenamefont {Xu}\ \emph {et~al.}(2021)\citenamefont {Xu},
  \citenamefont {Steinberg}, \citenamefont {Nguyen},\ and\ \citenamefont
  {GÃŒhne}}]{xu2021nogo}%
  \BibitemOpen
  \bibfield  {author} {\bibinfo {author} {\bibfnamefont {Z.-P.}\ \bibnamefont
  {Xu}}, \bibinfo {author} {\bibfnamefont {J.}~\bibnamefont {Steinberg}},
  \bibinfo {author} {\bibfnamefont {H.~C.}\ \bibnamefont {Nguyen}},\ and\
  \bibinfo {author} {\bibfnamefont {O.}~\bibnamefont {Gühne}},\ }\href {https://doi.org/10.48550/arXiv.2111.15010}
  {\bibinfo {title} {No-go theorem based on incomplete information of Wigner
  about his friend}} (\bibinfo {year} {2021}),\ {arXiv:2111.15010 [quant-ph]} \BibitemShut
  {NoStop}%
\bibitem [{\citenamefont {Nurgalieva}\ \emph {et~al.}(2018)\citenamefont {Nurgalieva}\ and\ \citenamefont
  {del Rio}}]{Nurgalieva2018}%
  \BibitemOpen
  \bibfield  {author} {\bibinfo {author} {\bibfnamefont {Nuriya}~\bibnamefont {Nurgalieva}}\ and\
  \bibinfo {author} {\bibfnamefont {L{\'i}dia}~\bibnamefont {del Rio}},\ }\href {https://doi.org/10.4204/EPTCS.287.16}
  {\bibinfo {title} {Inadequacy of Modal Logic in Quantum Settings}} (\bibinfo {year} {2018}),\ {arXiv:1804.01106 [quant-ph]} \BibitemShut
  {NoStop}%
\bibitem [{\citenamefont {Baumann}\ \emph {et~al.}(2021)\citenamefont {Baumann},
  \citenamefont {Del Santo}, \citenamefont {R. H. Smith}, \citenamefont {Giacomini}, \citenamefont {Castro-Ruiz},\ and\ \citenamefont {Brukner}}]{Baumann2021}%
  \BibitemOpen
  \bibfield  {author} {\bibinfo {author} {\bibfnamefont {Veronika}~\bibnamefont {Baumann}}, \bibinfo {author} {\bibfnamefont {Flavio}~\bibnamefont
  {Del Santo}}, \bibinfo {author} {\bibfnamefont {Alexander}~\bibnamefont
  {R. H. Smith}}, \bibinfo {author} {\bibfnamefont {Flaminia}~\bibnamefont
  {Giacomini}}, \bibinfo {author} {\bibfnamefont {Esteban}~\bibnamefont
  {Castro-Ruiz}},\ and\ \bibinfo {author} {\bibfnamefont {Caslav}~\bibnamefont
  {Brukner}},\ }\bibfield  {title} {\bibinfo {title} {Generalized probability rules from a timeless formulation of Wigner’s friend scenarios},\ }\href
  {https://doi.org/10.22331/q-2021-08-16-524} {\bibfield  {journal} {\bibinfo
  {journal} {Quantum}\ }\textbf {\bibinfo {volume} {5}},\ \bibinfo
  {pages} {594} (\bibinfo {year} {2021})}\BibitemShut {NoStop}%
\bibitem [{\citenamefont {Bell}(1964)}]{bell1964einstein}%
  \BibitemOpen
  \bibfield  {author} {\bibinfo {author} {\bibfnamefont {J.~S.}\ \bibnamefont
  {Bell}},\ }\bibfield  {title} {\bibinfo {title} {On the einstein podolsky
  rosen paradox},\ }\href {https://doi.org/10.1103/PhysicsPhysiqueFizika.1.195} {\bibfield  {journal} {\bibinfo  {journal}
  {Physics Physique Fizika}\ }\textbf {\bibinfo {volume} {1}},\ \bibinfo
  {pages} {195} (\bibinfo {year} {1964})}\BibitemShut {NoStop}%
\bibitem [{\citenamefont {Elitzur}\ \emph {et~al.}(1992)\citenamefont
  {Elitzur}, \citenamefont {Popescu},\ and\ \citenamefont
  {Rohrlich}}]{elitzur1992quantum}%
  \BibitemOpen
  \bibfield  {author} {\bibinfo {author} {\bibfnamefont {A.~C.}\ \bibnamefont
  {Elitzur}}, \bibinfo {author} {\bibfnamefont {S.}~\bibnamefont {Popescu}},\
  and\ \bibinfo {author} {\bibfnamefont {D.}~\bibnamefont {Rohrlich}},\
  }\bibfield  {title} {\bibinfo {title} {Quantum nonlocality for each pair in
  an ensemble},\ }\href {https://doi.org/10.1016/0375-9601(92)90952-I} {\bibfield  {journal} {\bibinfo  {journal}
  {Physics Letters A}\ }\textbf {\bibinfo {volume} {162}},\ \bibinfo {pages}
  {25} (\bibinfo {year} {1992})}\BibitemShut {NoStop}%
\bibitem [{\citenamefont {Braunstein}\ and\ \citenamefont
  {Caves}(1990)}]{braunstein1990wringing}%
  \BibitemOpen
  \bibfield  {author} {\bibinfo {author} {\bibfnamefont {S.~L.}\ \bibnamefont
  {Braunstein}}\ and\ \bibinfo {author} {\bibfnamefont {C.~M.}\ \bibnamefont
  {Caves}},\ }\bibfield  {title} {\bibinfo {title} {Wringing out better bell
  inequalities},\ }\href {https://doi.org/10.1016/0003-4916(90)90339-P} {\bibfield  {journal} {\bibinfo  {journal}
  {Annals of Physics}\ }\textbf {\bibinfo {volume} {202}},\ \bibinfo {pages}
  {22} (\bibinfo {year} {1990})}\BibitemShut {NoStop}%
\bibitem [{\citenamefont {Fine}(1982)}]{fine1982hidden}%
  \BibitemOpen
  \bibfield  {author} {\bibinfo {author} {\bibfnamefont {A.}~\bibnamefont
  {Fine}},\ }\bibfield  {title} {\bibinfo {title} {Hidden variables, joint
  probability, and the bell inequalities},\ }\href {https://doi.org/10.1103/PhysRevLett.48.291} {\bibfield
  {journal} {\bibinfo  {journal} {Physical Review Letters}\ }\textbf {\bibinfo
  {volume} {48}},\ \bibinfo {pages} {291} (\bibinfo {year} {1982})}\BibitemShut
  {NoStop}%
\bibitem [{\citenamefont {Hall}(2010{\natexlab{a}})}]{hall2010local}%
  \BibitemOpen
  \bibfield  {author} {\bibinfo {author} {\bibfnamefont {M.~J.}\ \bibnamefont
  {Hall}},\ }\bibfield  {title} {\bibinfo {title} {Local deterministic model of
  singlet state correlations based on relaxing measurement independence},\
  }\href {https://doi.org/10.1103/PhysRevLett.105.250404} {\bibfield  {journal} {\bibinfo  {journal} {Physical review
  letters}\ }\textbf {\bibinfo {volume} {105}},\ \bibinfo {pages} {250404}
  (\bibinfo {year} {2010}{\natexlab{a}})}\BibitemShut {NoStop}%
\bibitem [{\citenamefont {Chaves}\ \emph {et~al.}(2015)\citenamefont {Chaves},
  \citenamefont {Kueng}, \citenamefont {Brask},\ and\ \citenamefont
  {Gross}}]{chaves2015unifying}%
  \BibitemOpen
  \bibfield  {author} {\bibinfo {author} {\bibfnamefont {R.}~\bibnamefont
  {Chaves}}, \bibinfo {author} {\bibfnamefont {R.}~\bibnamefont {Kueng}},
  \bibinfo {author} {\bibfnamefont {J.~B.}\ \bibnamefont {Brask}},\ and\
  \bibinfo {author} {\bibfnamefont {D.}~\bibnamefont {Gross}},\ }\bibfield
  {title} {\bibinfo {title} {Unifying framework for relaxations of the causal
  assumptions in bell's theorem},\ }\href
  {https://doi.org/10.1103/PhysRevLett.114.140403} {\bibfield  {journal}
  {\bibinfo  {journal} {Phys. Rev. Lett.}\ }\textbf {\bibinfo {volume} {114}},\
  \bibinfo {pages} {140403} (\bibinfo {year} {2015})}\BibitemShut {NoStop}%
\bibitem [{\citenamefont {Hall}\ and\ \citenamefont
  {Branciard}(2020)}]{hall2020measurement}%
  \BibitemOpen
  \bibfield  {author} {\bibinfo {author} {\bibfnamefont {M.~J.}\ \bibnamefont
  {Hall}}\ and\ \bibinfo {author} {\bibfnamefont {C.}~\bibnamefont
  {Branciard}},\ }\bibfield  {title} {\bibinfo {title} {Measurement-dependence
  cost for bell nonlocality: Causal versus retrocausal models},\ }\href {https://doi.org/10.1103/PhysRevA.102.052228}
  {\bibfield  {journal} {\bibinfo  {journal} {Physical Review A}\ }\textbf
  {\bibinfo {volume} {102}},\ \bibinfo {pages} {052228} (\bibinfo {year}
  {2020})}\BibitemShut {NoStop}%
\bibitem [{\citenamefont {Chaves}\ \emph {et~al.}(2021)\citenamefont {Chaves},
  \citenamefont {Moreno}, \citenamefont {Polino}, \citenamefont {Poderini},
  \citenamefont {Agresti}, \citenamefont {Suprano}, \citenamefont {Barros},
  \citenamefont {Carvacho}, \citenamefont {Wolfe}, \citenamefont {Canabarro},
  \citenamefont {Spekkens},\ and\ \citenamefont
  {Sciarrino}}]{chaves2021causal}%
  \BibitemOpen
  \bibfield  {author} {\bibinfo {author} {\bibfnamefont {R.}~\bibnamefont
  {Chaves}}, \bibinfo {author} {\bibfnamefont {G.}~\bibnamefont {Moreno}},
  \bibinfo {author} {\bibfnamefont {E.}~\bibnamefont {Polino}}, \bibinfo
  {author} {\bibfnamefont {D.}~\bibnamefont {Poderini}}, \bibinfo {author}
  {\bibfnamefont {I.}~\bibnamefont {Agresti}}, \bibinfo {author} {\bibfnamefont
  {A.}~\bibnamefont {Suprano}}, \bibinfo {author} {\bibfnamefont {M.~R.}\
  \bibnamefont {Barros}}, \bibinfo {author} {\bibfnamefont {G.}~\bibnamefont
  {Carvacho}}, \bibinfo {author} {\bibfnamefont {E.}~\bibnamefont {Wolfe}},
  \bibinfo {author} {\bibfnamefont {A.}~\bibnamefont {Canabarro}}, \bibinfo
  {author} {\bibfnamefont {R.~W.}\ \bibnamefont {Spekkens}},\ and\ \bibinfo
  {author} {\bibfnamefont {F.}~\bibnamefont {Sciarrino}},\ }\bibfield  {title}
  {\bibinfo {title} {Causal networks and freedom of choice in bell's theorem},\
  }\href {https://doi.org/10.1103/PRXQuantum.2.040323} {\bibfield  {journal}
  {\bibinfo  {journal} {PRX Quantum}\ }\textbf {\bibinfo {volume} {2}},\
  \bibinfo {pages} {040323} (\bibinfo {year} {2021})}\BibitemShut {NoStop}%
\bibitem [{\citenamefont {Popescu}\ and\ \citenamefont
  {Rohrlich}(1994)}]{popescu1994quantum}%
  \BibitemOpen
  \bibfield  {author} {\bibinfo {author} {\bibfnamefont {S.}~\bibnamefont
  {Popescu}}\ and\ \bibinfo {author} {\bibfnamefont {D.}~\bibnamefont
  {Rohrlich}},\ }\bibfield  {title} {\bibinfo {title} {Quantum nonlocality as
  an axiom},\ }\href {https://doi.org/10.1007/BF02058098} {\bibfield  {journal} {\bibinfo  {journal}
  {Foundations of Physics}\ }\textbf {\bibinfo {volume} {24}},\ \bibinfo
  {pages} {379} (\bibinfo {year} {1994})}\BibitemShut {NoStop}%
\bibitem [{\citenamefont {Fitzi}\ \emph {et~al.}(2010)\citenamefont {Fitzi},
  \citenamefont {H{\"a}nggi}, \citenamefont {Scarani},\ and\ \citenamefont
  {Wolf}}]{fitzi2010non}%
  \BibitemOpen
  \bibfield  {author} {\bibinfo {author} {\bibfnamefont {M.}~\bibnamefont
  {Fitzi}}, \bibinfo {author} {\bibfnamefont {E.}~\bibnamefont {H{\"a}nggi}},
  \bibinfo {author} {\bibfnamefont {V.}~\bibnamefont {Scarani}},\ and\ \bibinfo
  {author} {\bibfnamefont {S.}~\bibnamefont {Wolf}},\ }\bibfield  {title}
  {\bibinfo {title} {The non-locality of n noisy popescu--rohrlich boxes},\
  }\href {https://doi.org/10.1088/1751-8113/43/46/465305} {\bibfield  {journal} {\bibinfo  {journal} {Journal of Physics
  A: Mathematical and Theoretical}\ }\textbf {\bibinfo {volume} {43}},\
  \bibinfo {pages} {465305} (\bibinfo {year} {2010})}\BibitemShut {NoStop}%
\bibitem [{\citenamefont {Mermin}(1990)}]{Mermin1990}%
  \BibitemOpen
  \bibfield  {author} {\bibinfo {author} {\bibfnamefont {N.~D.}\ \bibnamefont
  {Mermin}},\ }\bibfield  {title} {\bibinfo {title} {Extreme quantum
  entanglement in a superposition of macroscopically distinct states},\ }\href
  {https://doi.org/10.1103/PhysRevLett.65.1838} {\bibfield  {journal} {\bibinfo
  {journal} {Phys. Rev. Lett.}\ }\textbf {\bibinfo {volume} {65}},\ \bibinfo
  {pages} {1838} (\bibinfo {year} {1990})}\BibitemShut {NoStop}%
\bibitem [{\citenamefont {Brunner}\ \emph {et~al.}(2014)\citenamefont
  {Brunner}, \citenamefont {Cavalcanti}, \citenamefont {Pironio}, \citenamefont
  {Scarani},\ and\ \citenamefont {Wehner}}]{BrunnerEtAl14}%
  \BibitemOpen
  \bibfield  {author} {\bibinfo {author} {\bibfnamefont {N.}~\bibnamefont
  {Brunner}}, \bibinfo {author} {\bibfnamefont {D.}~\bibnamefont {Cavalcanti}},
  \bibinfo {author} {\bibfnamefont {S.}~\bibnamefont {Pironio}}, \bibinfo
  {author} {\bibfnamefont {V.}~\bibnamefont {Scarani}},\ and\ \bibinfo {author}
  {\bibfnamefont {S.}~\bibnamefont {Wehner}},\ }\bibfield  {title} {\bibinfo
  {title} {Bell nonlocality},\ }\href
  {https://doi.org/10.1103/RevModPhys.86.419} {\bibfield  {journal} {\bibinfo
  {journal} {Reviews of Modern Physics}\ }\textbf {\bibinfo {volume} {86}},\
  \bibinfo {pages} {419--478} (\bibinfo {year} {2014})}\BibitemShut {NoStop}%
\bibitem [{\citenamefont {Hall}(2010{\natexlab{b}})}]{Hall2010}%
  \BibitemOpen
  \bibfield  {author} {\bibinfo {author} {\bibfnamefont {M.~J.~W.}\
  \bibnamefont {Hall}},\ }\bibfield  {title} {\bibinfo {title} {Complementary
  contributions of indeterminism and signaling to quantum correlations},\
  }\href {https://doi.org/10.1103/PhysRevA.82.062117} {\bibfield  {journal}
  {\bibinfo  {journal} {Phys. Rev. A}\ }\textbf {\bibinfo {volume} {82}},\
  \bibinfo {pages} {062117} (\bibinfo {year} {2010}{\natexlab{b}})}\BibitemShut
  {NoStop}%
\bibitem [{\citenamefont {Wehner}(2006)}]{Wehner2006}%
  \BibitemOpen
  \bibfield  {author} {\bibinfo {author} {\bibfnamefont {S.}~\bibnamefont
  {Wehner}},\ }\bibfield  {title} {\bibinfo {title} {Tsirelson bounds for
  generalized clauser-horne-shimony-holt inequalities},\ }\href
  {https://doi.org/10.1103/PhysRevA.73.022110} {\bibfield  {journal} {\bibinfo
  {journal} {Phys. Rev. A}\ }\textbf {\bibinfo {volume} {73}},\ \bibinfo
  {pages} {022110} (\bibinfo {year} {2006})}\BibitemShut {NoStop}%
\bibitem [{\citenamefont {Einstein}\ \emph {et~al.}(1935)\citenamefont
  {Einstein}, \citenamefont {Podolsky},\ and\ \citenamefont
  {Rosen}}]{einstein1935can}%
  \BibitemOpen
  \bibfield  {author} {\bibinfo {author} {\bibfnamefont {A.}~\bibnamefont
  {Einstein}}, \bibinfo {author} {\bibfnamefont {B.}~\bibnamefont {Podolsky}},\
  and\ \bibinfo {author} {\bibfnamefont {N.}~\bibnamefont {Rosen}},\ }\bibfield
  {title} {\bibinfo {title} {Can quantum-mechanical description of physical
  reality be considered complete?},\ }\href {https://doi.org/10.1103/PhysRev.47.777} {\bibfield  {journal}
  {\bibinfo  {journal} {Physical review}\ }\textbf {\bibinfo {volume} {47}},\
  \bibinfo {pages} {777} (\bibinfo {year} {1935})}\BibitemShut {NoStop}%
\bibitem [{\citenamefont {De~Vicente}(2014)}]{de2014nonlocality}%
  \BibitemOpen
  \bibfield  {author} {\bibinfo {author} {\bibfnamefont {J.~I.}\ \bibnamefont
  {De~Vicente}},\ }\bibfield  {title} {\bibinfo {title} {On nonlocality as a
  resource theory and nonlocality measures},\ }\href {https://doi.org/10.1088/1751-8113/47/42/424017} {\bibfield
  {journal} {\bibinfo  {journal} {Journal of Physics A: Mathematical and
  Theoretical}\ }\textbf {\bibinfo {volume} {47}},\ \bibinfo {pages} {424017}
  (\bibinfo {year} {2014})}\BibitemShut {NoStop}%
\bibitem [{\citenamefont {Brito}\ \emph {et~al.}(2018)\citenamefont {Brito},
  \citenamefont {Amaral},\ and\ \citenamefont {Chaves}}]{brito2018quantifying}%
  \BibitemOpen
  \bibfield  {author} {\bibinfo {author} {\bibfnamefont {S.~G.~A.}\
  \bibnamefont {Brito}}, \bibinfo {author} {\bibfnamefont {B.}~\bibnamefont
  {Amaral}},\ and\ \bibinfo {author} {\bibfnamefont {R.}~\bibnamefont
  {Chaves}},\ }\bibfield  {title} {\bibinfo {title} {Quantifying bell
  nonlocality with the trace distance},\ }\href
  {https://doi.org/10.1103/PhysRevA.97.022111} {\bibfield  {journal} {\bibinfo
  {journal} {Phys. Rev. A}\ }\textbf {\bibinfo {volume} {97}},\ \bibinfo
  {pages} {022111} (\bibinfo {year} {2018})}\BibitemShut {NoStop}%
\bibitem [{\citenamefont {Wolfe}\ \emph {et~al.}(2020)\citenamefont {Wolfe},
  \citenamefont {Schmid}, \citenamefont {Sainz}, \citenamefont {Kunjwal},\ and\
  \citenamefont {Spekkens}}]{wolfe2020quantifying}%
  \BibitemOpen
  \bibfield  {author} {\bibinfo {author} {\bibfnamefont {E.}~\bibnamefont
  {Wolfe}}, \bibinfo {author} {\bibfnamefont {D.}~\bibnamefont {Schmid}},
  \bibinfo {author} {\bibfnamefont {A.~B.}\ \bibnamefont {Sainz}}, \bibinfo
  {author} {\bibfnamefont {R.}~\bibnamefont {Kunjwal}},\ and\ \bibinfo {author}
  {\bibfnamefont {R.~W.}\ \bibnamefont {Spekkens}},\ }\bibfield  {title}
  {\bibinfo {title} {Quantifying bell: The resource theory of nonclassicality
  of common-cause boxes},\ }\href {https://doi.org/10.22331/q-2020-06-08-280} {\bibfield  {journal} {\bibinfo
  {journal} {Quantum}\ }\textbf {\bibinfo {volume} {4}},\ \bibinfo {pages}
  {280} (\bibinfo {year} {2020})}\BibitemShut {NoStop}%
\bibitem [{\citenamefont {Brask}\ and\ \citenamefont
  {Chaves}(2017)}]{Brask_2017}%
  \BibitemOpen
  \bibfield  {author} {\bibinfo {author} {\bibfnamefont {J.~B.}\ \bibnamefont
  {Brask}}\ and\ \bibinfo {author} {\bibfnamefont {R.}~\bibnamefont {Chaves}},\
  }\bibfield  {title} {\bibinfo {title} {Bell scenarios with communication},\
  }\href {https://doi.org/10.1088/1751-8121/aa5840} {\bibfield  {journal}
  {\bibinfo  {journal} {Journal of Physics A: Mathematical and Theoretical}\
  }\textbf {\bibinfo {volume} {50}},\ \bibinfo {pages} {094001} (\bibinfo
  {year} {2017})}\BibitemShut {NoStop}%
\bibitem [{\citenamefont {{\v{S}}upi{\'{c}}}\ \emph {et~al.}(2016)\citenamefont {{\v{S}}upi{\'{c}}},
  \citenamefont {Augusiak}, \citenamefont {Salavrakos}\ and\ \citenamefont {Ac{\'i}n}}]{Supic2016}%
  \BibitemOpen
  \bibfield  {author} {\bibinfo {author} {\bibfnamefont {I.}~\bibnamefont
  {{\v{S}}upi{\'{c}}}}, \bibinfo {author} {\bibfnamefont {R.}~\bibnamefont {Augusiak}}, \bibinfo {author} {\bibfnamefont {A.}~\bibnamefont {Salavrakos}}\ and\ \bibinfo {author} {\bibfnamefont {A.}~\bibnamefont {Ac{\'i}n}},\
  }\bibfield  {title} {\bibinfo {title} {Self-testing protocols based on the chained Bell inequalities},\
  }\href {https://doi.org/10.1088/1367-2630/18/3/035013} {\bibfield  {journal}
  {\bibinfo  {journal} {New Journal of Physics}\
  }\textbf {\bibinfo {volume} {18}},\ \bibinfo {pages} {035013} (\bibinfo
  {year} {2016})}\BibitemShut {NoStop}%
\end{thebibliography}
\end{document}